\documentclass[journal]{IEEEtran}
\IEEEoverridecommandlockouts

\usepackage{times}
\usepackage{epsfig}
\usepackage{amsmath}
\usepackage{amssymb}
\usepackage{multicol}
\usepackage{soul}

\usepackage{subfigure}
\usepackage{multirow}
\usepackage{xcolor}
\usepackage{pifont}
\newcommand{\cmark}{\ding{51}}
\newcommand{\xmark}{\ding{53}}

\usepackage[boldmath]{numprint}

\usepackage[breaklinks=true,bookmarks=false]{hyperref}

\title{In-Orbit Lunar Satellite Image Super Resolution \\for Selective Data Transmission}
\author{
	\IEEEauthorblockN{Atal Tewari\IEEEauthorrefmark{2}$^{,\star}$, Chennuri Prateek\IEEEauthorrefmark{2}$^{,\star}$, Nitin Khanna\IEEEauthorrefmark{3}$^{,+}$
    \thanks{Codes and dataset related to this paper can be downloaded using   	\href{https://docs.google.com/forms/d/e/1FAIpQLSfN8FB-y8wbK4RlgCmco8d6X1nBRxBGY0jDBcQfyKE_JQ2sZQ/viewform}{this link}.
    	This research is based upon work partially supported by the ISRO, Department of Space, Government of India under the Award number ISRO/SSPO/Ch-1/2016-17.  Atal Tewari is supported by TCS Research Scholarship. Any opinions, findings, and conclusions or recommendations expressed in this material are those of the author(s) and do not necessarily reflect the views of the funding agencies. 
		We also acknowledge the use of data from NASA's LRO spacecraft, which was downloaded from the archives of the USGS.  
	}
	\thanks{$^{+}$Corresponding author. Please address all correspondences to Nitin Khanna, Multimedia Analysis and Security (MANAS) Lab, Electrical Engineering and Computer Science, Indian Institute of Technology Bhilai, India. E-mail address: nitin@iitbhilai.ac.in}
	\thanks{$^\star$Equal contribution}
}
\\
	\IEEEauthorblockA{\IEEEauthorrefmark{2}Electrical Engineering, Indian Institute of Technology Gandhinagar, India}\\
	\IEEEauthorblockA{\IEEEauthorrefmark{3} Electrical Engineering and Computer Science, Indian Institute of Technology Bhilai, India}
}

\begin{document}

\maketitle

\begin{abstract}
Rapid technological advancements have tremendously increased the data acquisition capabilities of remote sensing satellites. 
However, the data utilization efficiency in satellite missions is very low. 
This growing data also escalates the cost required for data downlink transmission and post-processing. Selective data transmission based on in-orbit inferences will address these issues to a great extent. 
Therefore, to decrease the cost of the satellite mission, we propose a novel system design for selective data transmission, based on in-orbit inferences. 
As the resolution of images plays a critical role in making precise inferences, we also include in-orbit super-resolution (SR) in the system design. 
We introduce a new image reconstruction technique and a unique loss function to enable the execution of the SR model on low-power devices suitable for satellite environments. 
We present a residual dense non-local attention network (RDNLA) that provides enhanced super-resolution outputs to improve the SR performance. 
SR experiments on Kaguya digital ortho maps (DOMs) demonstrate that the proposed SR algorithm outperforms the residual dense network (RDN) in terms of PSNR and block-sensitive PSNR by a margin of $+0.1$ dB and $+0.19$ dB, respectively. 
The proposed SR system consumes 48\% less memory and 67\% less peak instantaneous power than the standard SR model, RDN, making it more suitable for execution on a low-powered device platform. 
\end{abstract}

\section{Introduction}
The extraction of high resolution (HR) imagery from one of its low resolution (LR) counterparts is a topic of great significance in the field of computer vision. This task, commonly referred to as single image super-resolution (SISR), is an ill-posed problem since there are always multiple HR images corresponding to a single LR image~\cite{Arefin_2020_CVPR_Workshops}.
SISR has a wide range of computer vision and image processing applications such as medical diagnostics, satellite imaging, and other high-level vision tasks~\cite{Arefin_2020_CVPR_Workshops}. 
Specifically, in lunar satellite imaging, image resolution plays a critical role in studies such as topological and geomorphological analysis, elemental/chemical composition, and mineralogy to understand the moon's origin and evolution. 

For all the satellite subsystems, the sun acts as the primary energy source, whose energy is transferred from electromagnetic to electric using solar panels mounted around the satellite.
Due to solar energy's renewability, there is enough energy available for the satellite mission. However,  there is a hard limitation on the power available due to the hardware constraints of the solar panels. 
Therefore, low powered devices are preferred for satellite subsystems to utilize power more effectively. 

Continuous improvements in satellite imaging technology in terms of spatial and spectral resolution lead to an exponential increase in satellite data. The satellite's strong data acquisition capacity brings tremendous pressure on the satellite data downlink transmission and post-processing~\cite{bing2011intelligent}. Moreover, there is a vast gap between the amount of data received and the amount in actual use, implying extremely low data utilization efficiency~\cite{bing2011intelligent}. A viable solution for such difficulties is to design a selective data transmission technique based on in-orbit satellite inferences. As image resolution plays a crucial role in making accurate inferences and high-resolution sensors are generally expensive, there is a necessity to include a super-resolution algorithm (executable on low-powered devices) in the selective data transmission system that can extract high-resolution (HR) images from the satellite's low-resolution sensors. Therefore, there is a need for an in-orbit satellite super-resolution algorithm executable on low-powered devices suitable for satellite environments. 
\begin{figure*}[!htb]
    \centering
    \includegraphics[width=0.9\textwidth]{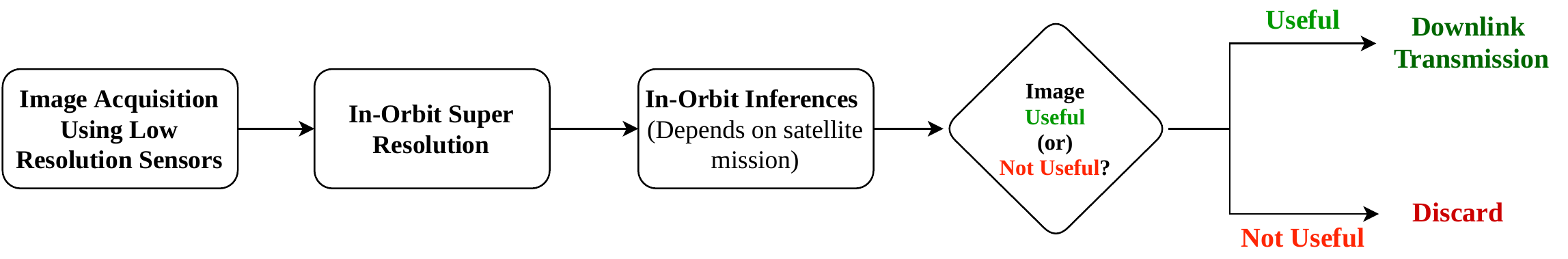}
    \caption{Overview of Selective Data Transmission System Design}
    \label{fig:flowchart_selective_data_transmission}
\end{figure*}
With the rapid advancement of deep learning (DL) technology in recent years, deep learning-based SR models have been actively explored, among which single image super-resolution (SISR) has become the mainstream. 
Lukas et al.~\cite{liebel2016single} demonstrated one of the pioneer works, which signified that deep-learning (DL) was an effective solution for remote sensing image SR. Starting from LGCNet~\cite{7937881} to the latest CNN-based SISR architectures like  MHAN~\cite{9151234}, SMSR~\cite{SMSR}, SR networks for remote sensing have evolved dramatically in recent times. However, most remote sensing SISR methods consume significantly high peak instantaneous power thus cannot be utilized as a part of the in-orbit SR algorithm. Though few SR DL models~\cite{kim2018real,lee2019mobisr,liu2021splitsr,cheng2021mfagan}, can be executed on low-powered devices, they experience a significant trade-off in the output quality. Therefore, utilizing such algorithms would lead to inaccurate in-orbit inferences, thus, decreasing the overall performance of the selective data transmission technique.

To address the drawbacks mentioned in the previous paragraphs, we propose a novel system design for selective data transmission that includes an in-orbit SR step to make accurate in-orbit inferences. For the SR algorithm to be executable on low-powered devices, we follow a simple yet effective SR evaluation procedure as shown in Figure~\ref{fig:framework_reconstruction}. We introduce a unique overlap-reconstruction mechanism that extracts patch regions with more contextual information for efficient SR image reconstruction using output patches. Along the lines of the overlap reconstruction approach, we introduce a novel and efficient loss function, mask-PSNR, that allows the model to produce enhanced results by laying higher emphasis on the regions of interest. 

To improve the SR performance, we modify the residual dense network~\cite{Zhang_2020_TPAMI} into a residual dense non-local attention network that extracts the original LR image's hierarchical features by utilizing the residual, dense, and attention mechanism to capture long-range dependencies for enhanced reconstruction. Our RDNLA is built on our proposed residual dense non-local attention block (RDNLB) shown in Figure~\ref{fig:framework}. The novel and complex combination of residual, dense and non-local attention in RDNLB strengthen the model's long-range dependent feature propagation by re-utilizing the features in a feed-forward fashion~\cite{Densenet} which further enhances the feature representation ability, thus, improving the SR performance as a whole. A more detailed description of the proposed model can be found in Section~\ref{subsec:RDNLA}. In summary, our main contributions are as follows:
\begin{itemize}
    \item We identify an unsolved problem of low data utilization efficiency in satellite missions. We resolve the issue by introducing a novel system design for the selective data transmission based on in-orbit inferences.
    \item We propose an overlap reconstruction mechanism that extracts regions with adequate contextual information for the in-orbit super-resolution algorithm to be executable on low-powered devices suitable for satellite environments. Along the lines of overlap reconstruction, we present a new loss function, mask PSNR, that lets the SR DL model emphasize regions having rich contextual information.
    \item To improve the SR performance, we introduce a residual dense non-local attention network that utilizes an intricate combination of residual, dense and non-local attention blocks to effectively extract long-range dependent hierarchical features and provide enhanced SR output.
    \item To the best of our knowledge, this is the first work to present an SR algorithm lunar digital ortho maps (DOMs). Moreover, this is the first deep learning-based algorithm proposed for SR of lunar satellite data (Ex: DOMs, DEMs).
\end{itemize}

\section{Related Work}
In recent years, with the increase in demand for HR images in remote sensing and the deep learning-based methods broad applicability, many researchers use deep learning-based methods for SR of remote sensing images. Lukas et al.~\cite{liebel2016single} used convolution layers for SR of multi-spectral images. To learn a multi-level representation including local and global details, a local-global combined network (LGCNet) was proposed by Lei et al.~\cite{7937881}.  A multi-perception attention network (MPSR)~\cite{rs11232857} containing enhanced residual block (ERB), and residual channel attention group (RCAG) was developed to efficiently capture the prior information and adaptively focus on informative features. Pan et al.~\cite{8732688} utilized residual, dense along with recursive blocks to improve the performance of super-resolved remote sensing images. A recent paper~\cite{9151234} uses a novel higher-order attention (HOA) mechanism was introduced in which consists of a feature extraction network and a feature refinement network to extract the input image high-frequency details effectively. The above mentioned SR solutions provide enhanced results; however, they are computationally complex and memory expensive, therefore, not suitable for executing on a low-powered device. 

In recent years, there is significant research presented in the direction of low-powered device SR~\cite{kim2018real,lee2019mobisr,liu2021splitsr,cheng2021mfagan}. Though these approaches handle the computation complexity and memory constraints to a great extent, the performance is significantly compromised compared to the standard DL architectures. Therefore, these procedures can not be utilized for selective data transmission since it would decrease the system's overall effectiveness. Our work mainly focuses on proposing an SR algorithm that is executable on low-powered devices without compromising on the quality of the SR output. 

\section{Methodology}

\subsection{System Design for Selective Data Transmission}
\begin{figure*}[!htb]
    \centering
    \includegraphics[width=\textwidth]{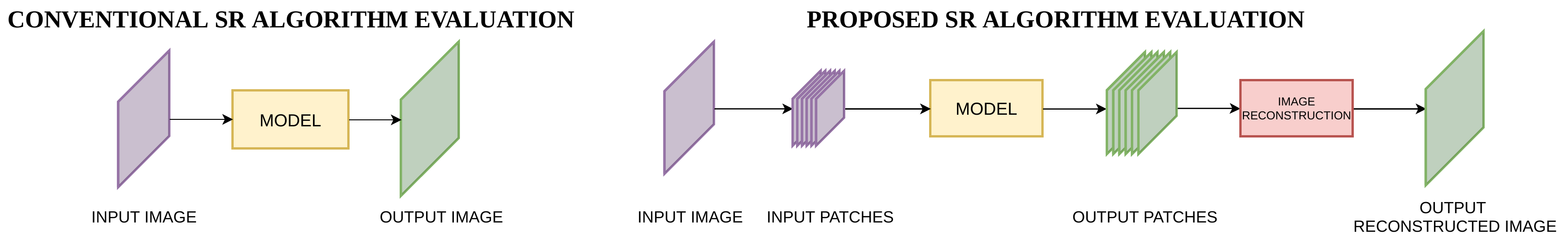}
    \caption{Comparison of proposed SR evaluation with conventional SR evaluation.}
    \label{fig:framework_reconstruction}
\end{figure*}
To address the problem of low data utilization efficiency in satellite missions, we present a new system design for selective data transmission based on in-orbit inferences as shown in Figure~\ref{fig:flowchart_selective_data_transmission}. The image acquisition is performed using low-resolution sensors since they are less expensive. The LR image is provided as an input to the in-orbit SR algorithm. For the SR algorithm to be executable on low-power devices, we follow a unique SR evaluation method as shown in Figure~\ref{fig:framework_reconstruction}. More information about the in-orbit SR algorithm can be found in Sections~\ref{subsec:RDNLA},~\ref{subsec:OR_MP}. The SR output generated through the in-orbit SR algorithm is utilized to make in-orbit inferences. The kind of inferences made entirely depends on the goals of the satellite mission; some of them include crater detection, mineral detection, and others. Experiments on in-orbit inferences are beyond the scope of this paper. Based on the results obtained through in-orbit inferences, a decision is made whether the image is useful or not, depending on which the image is transmitted or discarded. By utilizing such a pipeline, only the valuable images are downlink transmitted, increasing the data utilization efficiency and decreasing the overall cost of downlink transmission. 

\subsection{Residual Dense Non-Local Attention Network}
\label{subsec:RDNLA}
\begin{figure*}[!htb]
    \centering
    \includegraphics[width=\textwidth]{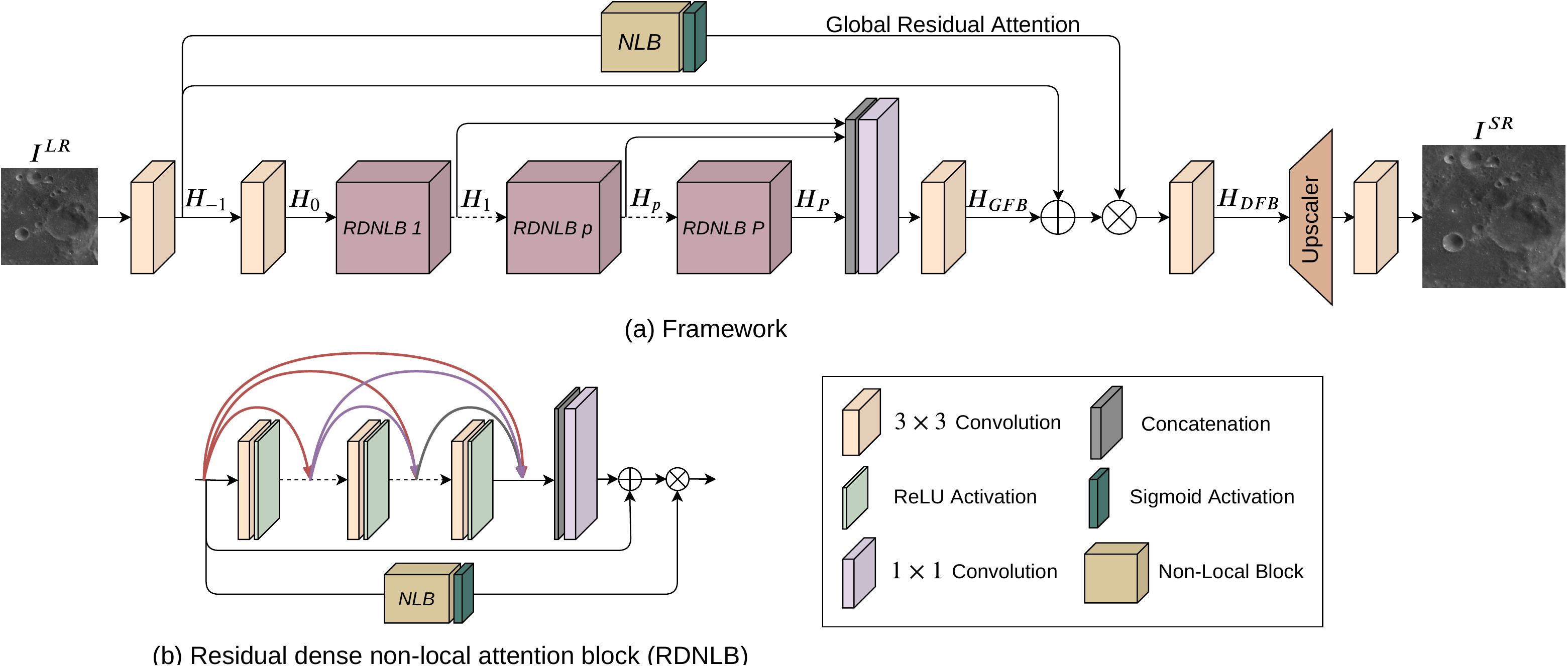}
    \caption{Overview of the proposed residual dense non-local attention network (RDNLA).}
    \label{fig:framework}
\end{figure*}
As shown in Figure~\ref{fig:framework}, our network can be broadly divided into four parts: shallow feature extraction (SFE), residual dense non-local attention blocks (RDNLBs), dense feature blending (DFB), and up-scaling (US). 
Let us assume $I^{LR}$ and $I^{SR}$ to be the low resolution input and super resolution output, respectively. 
SFE comprises of two $3\times3$ convolution layers that extract low-level features. 
Output of the second Conv layer is represented as
\begin{align} \label{eq:5}
    H_{0} = F_{SFE2}(H_{-1}) = F_{SFE2}(F_{SFE1}(I^{LR})),
\end{align}
where $F_{SFE1}$ and $F_{SFE2}$ represent the convolution operation of first and second convolution layers, respectively. $H_{0}$ is then used as an input to residual dense non-local attention blocks (RDNLBs) to extract dense long-range dependent hierarchical features. The output of the $p^{th}$ ($1 \leq p \leq P$) RDNLB can be represented as 

\begin{equation}\label{eq:5}
\begin{split}
H_{p} &= F_{RDNLB,p}(H_{p-1}) \\
      &=  F_{RDNLB,p}(F_{RDNLB,p-1}(...(F_{RDNLB,1}(H_{0}))...)),
\end{split}
\end{equation}

where $F_{RDNLB,p}$ indicates the composite functions such as convolution, ReLU~\cite{pmlr-v15-glorot11a}, dense connections~\cite{Densenet} of the $p^{th}$ RDNLB.
The hierarchical feature maps are further passed through the process of DFB that involves global feature blending (GFB) and global residual attention (GRA) mechanism. DFB blends the features of all the preceding RDNLBs and extracts global long-range dependent hierarchical features which are then used as input for up-scaling (US). DFB can be represented as, 

\begin{equation}
   H_{DFB} = F_{DFB}(H_{-1},H_{0},...,H_{P}), 
\end{equation}
where $F_{DFB}$ denotes the composite function of GFB and GRA.
Lastly, we utilize the local and global hierarchical features as input for US, followed by a $3\times3$ Conv layer for extracting the SR output.  

\textbf{Residual Dense Non-Local Attention Block.} Our proposed RDNLB is shown in Figure~\ref{fig:framework}. It consists of densely connected layers, local feature blending (LFB), and local residual attention (LRA) leading to a coupled memory (CM) mechanism. 

The coupled memory mechanism is achieved by transferring the previous RDNLB output to each layer of the current RDNLB. Mathematically, the output of the $d^{th}$ Conv layer of the $p^{th}$ RDNLB is 
\begin{align}\label{eq:2}
    H_{(p,d)} &= \Gamma(W_{(p,d)}[H_{p-1}, H_{(p,1)},..., H_{(p,d-1)}]),
\end{align} where $H_{p-1}$ and $H_{p}$ are the input and output of the $p^{th}$ RDNLB, each consisting of $G_b$ feature maps. $W_{(p,d)}$ is the weight vector of the $d^{th}$ Conv layer (the bias term is excluded for simplicity). 
$\Gamma$ represents the ReLU~\cite{pmlr-v15-glorot11a} activation function. 
$[H_{p-1}, H_{(p,1)},..., H_{(p,d-1)}]$ represents the concatenation of channel outputs produced by the $(p-1)^{th}$ RDNLB and the convolutional layers $1,..., (d-1)$ in the $p^{th}$ RDNLB. 

Local feature blending is the process of adaptively fusing the outputs of preceding RDNLB and the current RDNLB's Conv layers. The concatenation operation itself is not viable since the size of feature-maps would change for every RDNLB. Moreover, as the growth rate $G$ increases, the network's depth also increases, resulting in training difficulty. Therefore, to ease the training difficulty, similar to Memnet~\cite{Tai_2017_ICCV}, RDN~\cite{Zhang_2020_TPAMI}, we utilize $1\times1$ convolutional layer to control the channel output information adaptively. 
LFB can be formulated as
\begin{align}\label{eq:3}
    H^{p}_{LF} & = W^{p}_{1\times1}([H_{p-1}, H_{(p,1)},...,H_{(p,D-1)}, H_{(p,D)}]),
\end{align} 
where $H^{p}_{LF}$ represents the LFB output, $W^{p}_{1\times1}$ denotes the  $1\times1$ convolution layer operation of the $p^{th}$ RDNLB. 

Local residual attention is a combination of local residual learning and non-local attention mechanism. As there are several convolutional layers in a single RDNLB, local residual learning plays a critical role in enhancing the information flow throughout the network. The local residual learning output can be represented as 
\begin{equation}
    H^p_{LR} = H^p_{LF} + H_{p-1}.
\end{equation}

The main task of non-local attention mechanism, consisting of non-local attention~\cite{wang2018non} block (NLB) followed by sigmoid activation, is to grasp information of larger receptive field size, so that it is feasible to obtain sophisticated long-range dependent hierarchical features. 
The final output of the $d^{th}$ RDNLB can be represented as 
\begin{equation}
    H_p = H^p_{LR}\, \Lambda(F_{NLB}(H_{p-1})),
\end{equation}
where $F_{NLB}$ and $\Lambda$ represents the non-local attention operation and sigmoid activation.

\textbf{Dense Feature Blending.} After extracting the local dense long-range dependent hierarchical features using RDNLBs, we further introduce DFB to exploit the features globally. Our DFB consists of GFB and GRA mechanism. 

Global feature blending brings out the global hierarchical features by fusing the states of all the RDNLBs. 
GFB can be represented as
\begin{align}\label{eq:7}
    H_{GFB} &= F_{GFB}([H_1,..., H_{P-1},H_P]),
\end{align} 
where $[H_1,..., H_{P-1},H_P]$ represents the concatenation of $1,\dotsc,P$ RDNLB outputs. $F_{GFB}$ indicates the composite function of $1\times1$ and $3\times3$ convolutional layers. The $1\times1$ convolutional layer adaptively controls the output information and the $3\times3$ convolution modifies and extracts the global hierarchical features which are further introduced into the GRA mechanism. 

Global residual attention consists of global residual learning followed by the non-local attention mechanism. The global residual learning improves the information flow in the network while mitigating the vanishing gradient problem. 
The non-local attention mechanism is introduced to drive the network to extract long-range dependent hierarchical features globally. This encourages the model to extract details of larger receptive field, thus producing long-range dependent feature maps for improved image SR. The output of dense feature blending that serves as an input to the up-scaling block is 
\begin{align} \label{eq:8}
    H_{DFB} &= F_{GRA}(H_{-1}, H_{GFB}),
\end{align}
where $F_{GRA}$ denotes the composite function of global residual learning, non-local attention, and $3\times3$ convolution. 
\begin{figure*}[!htb]
    \centering
    \includegraphics[width=\linewidth]{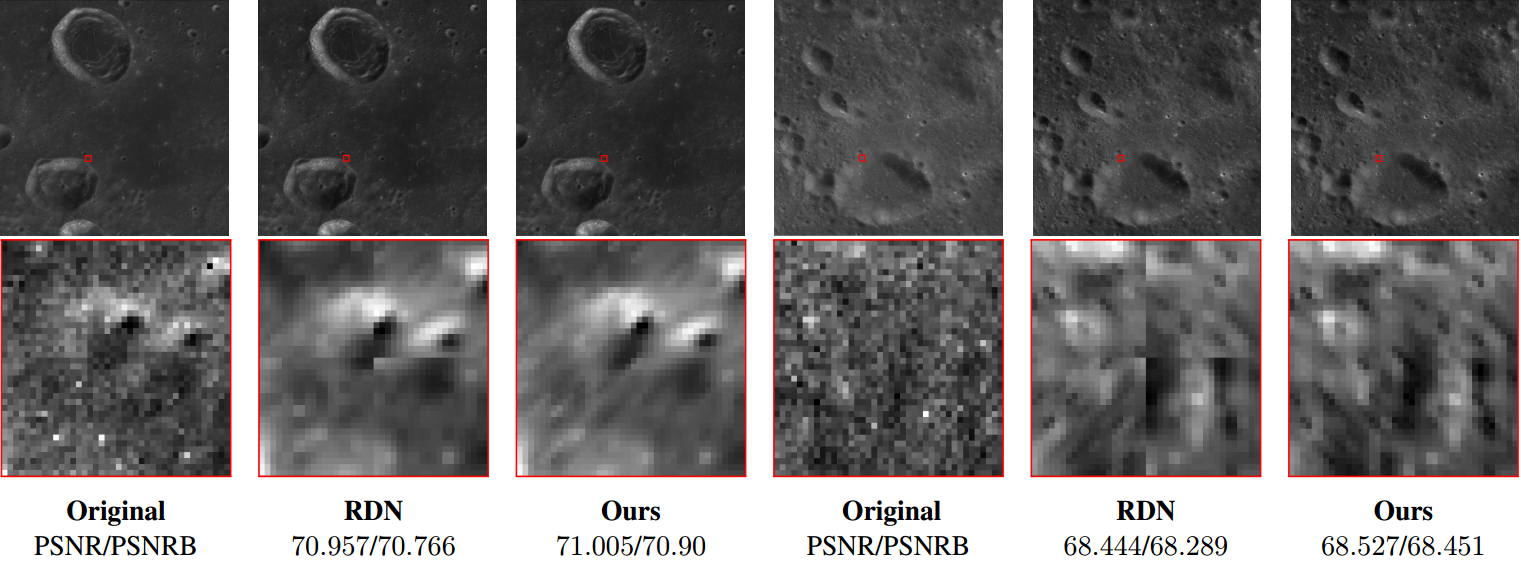}
    \caption{Visual SR results of the baseline (RDN) and our proposed (RDNLA) models trained on Kaguya DOMs~\cite{haruyama2008global} ($\text{x2}$).}
    \label{fig:OR_BA}
\end{figure*}

\subsection{Overlap Reconstruction and Mask-PSNR}
\label{subsec:OR_MP}
In a low-powered device, the input size plays a critical role in the amount of multiply and accumulate operations and memory access operations~\cite{chen2020deep}.
To be executable on such devices, we follow a patch-based prediction approach for extracting the SR outputs as shown in Figure~\ref{fig:framework_reconstruction}. During image reconstruction, non-overlapping stacking of patches leads to significant blocking artifacts in the reconstructed output, as shown in Figure~\ref{fig:OR_BA}. It may be due to the lack of adequate contextual information at the boundaries of the patch leading to ineffective extraction of hierarchical features.
\begin{figure*}[!htb]
    \centering
    \includegraphics[width=0.55\linewidth]{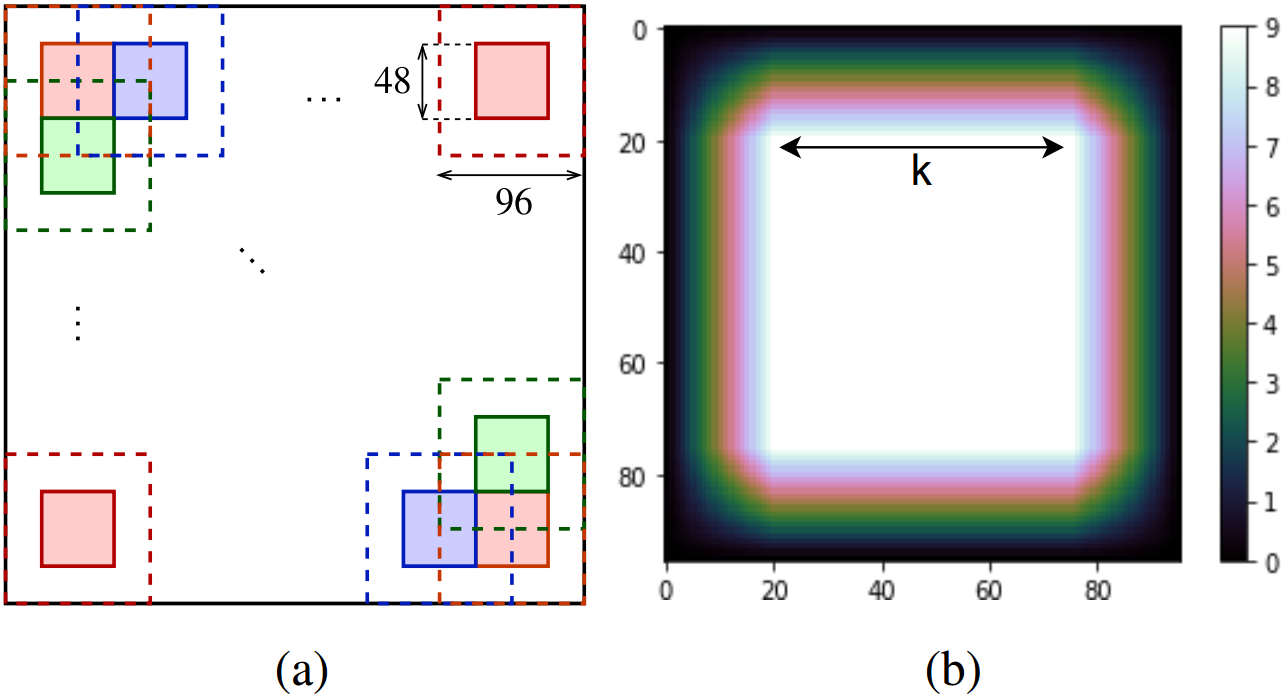}
     \caption{(a) Overlap reconstruction.  The effective patches extracted of size $48\times48$ are represented in red, green and blue. (b) "2-D box linear decay" mask.}
    \label{fig:OR_MASK}
\end{figure*}
To alleviate this problem, we propose an overlap reconstruction approach that provides $50\%$ overlapping patches as input to the network. As shown in Figure~\ref{fig:OR_MASK}, the reconstruction phase extracts only the central region of each patch ($25\%$ of the patch) that has enough contextual information for effective feature extraction.

We also prefer our model to offer increased attention at the patch's central region for superior results. 
Therefore, we introduce mask-PSNR, a novel loss function that utilizes a mask ($M$) to let the network provide additional attention to the patch's central region. 
As shown in the Figure~\ref{fig:OR_MASK}, $k$ depicts the size of the region that the model emphasizes the most. Mathematically, our loss function can be expressed as
\begin{align}\label{eq:13}
    \text{Mask-PSNR} = 10\,log_{10}\:\frac{N^2I^2_{max}}{\sum_{u,v} M(u,v)\left[I^{HR} (u,v) - I^{SR} (u,v)\right]^2},
\end{align}
where $I_{max}$, $N^2$ represent the maximum pixel intensity and size of the patch. $M(u,v)$, $I^{HR}(u,v)$, and $I^{SR}(u,v)$ represent the value of mask, HR patch, and SR patch, at location $(u,v)$.
We term the type of mask used in our work as ``2-D box linear decay". 

\subsection{Implementation Details}
In our proposed algorithm, we utilize zero padding to maintain the output size of every convolution layer constant. Every Conv operation in the RDNLBs has $G$ filters; however, shallow feature extraction, local, and global feature blending Conv layers have $G_{b} = 64$ filters. For up-scaling (US), we utilize the transpose convolution layer followed by a Conv layer with a single output channel as we output grayscale images. However, the network can also be utilized for processing color images.

\section{Settings}
\textbf{Dataset, Degradation Model, and Metrics: } 
The digital ortho maps (DOMs) utilized in this work are captured from Kaguya terrain camera~\cite{haruyama2008global},  span a range of  [$-63^{\circ}$, $63^{\circ}$] in latitude and [$30^{\circ}$, $33^{\circ}$] in longitude.  
We generate both the input LR ($100$ m/pixel) and ground truth HR images ($50$ m/pixel) by applying the bicubic interpolation on the original DOMs ($\sim$7.4 m/pixel). 
Total images considered for training and testing are 120 and 21, respectively. 
The HR and LR images have a size of $1820\times1820$ and $910\times910$. 
The SR results are evaluated using PSNR and PSNRB~\cite{psnr-b}.

\textbf{Training Settings: } 
We incorporate the following settings for lunar DOMs. 
We train our model for $75$ epochs with a learning rate of $10^{-4}$ and utilized Adam optimizer~\cite{adam} to update the weights. 
Each LR and HR image is divided into overlapping patches of size $48 \times 48$ and $96 \times 96$, respectively, with an overlap region of 50\%. 

\section{Results}

\begin{table*}[ht!]
    \centering
    \caption{SR results based on Kaguya DOMs~\cite{haruyama2008global}. Model parameters include $\text{P}=16$, $\text{D}=6$, and $\text{G}=32$. The best and baseline results are \textbf{highlighted} and \underline{underlined}.}
    \label{tab:main_results}
    \resizebox{0.8\textwidth}{!}
    {
    \begin{tabular}{|c c c c c |c c c c|}
        \hline
        \multicolumn{2}{|c}{Loss Type} &  \multicolumn{2}{c}{Upsampling} & Overlap  & \multicolumn{2}{c}{PSNR} & \multicolumn{2}{c|}{PSNR-B}\\ 
        L1 Loss& Mask-PSNR & Subpixel & DeConv & Reconstruction& RDN~\cite{Zhang_2020_TPAMI} & \textbf{Ours} & RDN~\cite{Zhang_2020_TPAMI} & \textbf{Ours} \\ \hline \hline
        \cmark &  & \cmark &  &  & $\underline{69.863}$ & $69.880$ & $\underline{69.690}$ & $69.714$\\ 
        \cmark &  & \cmark &  & \cmark & $69.919$ & $69.918$ & $69.841$ & $69.841$ \\ 
        \cmark &  &  & \cmark &  & $69.864$ & $69.877$ & $69.696$ & $69.709$\\ 
        \cmark &  &  & \cmark & \cmark & $69.904$ & $69.917$& $69.824$ & $69.840$\\ 
         &  \cmark & \cmark &  & \cmark & $69.925$ & $69.929$ & $69.844$ & $69.858$\\ 
         &  \cmark &  & \cmark & \cmark & $69.934$ & $\textbf{69.971}$ & $69.853$ & $\textbf{69.879}$\\ \hline
    \end{tabular}
    }
\end{table*}

We compare our network with one of the state-of-the-art natural image SR model, RDN~\cite{Zhang_2020_TPAMI}, trained on Kaguya DOMs. 
Table~\ref{tab:main_results} illustrates the quantitative comparisons for various settings between RDN and our model (RDNLA) for $\times2$ SR.
All the experiments in Table~\ref{tab:main_results} follow the patch-based prediction approach. However, experiments that do not follow overlap reconstruction utilize the non-overlapping stacking approach to reconstruct the SR output. 

The networks utilizing the L1 loss function depict the overlap reconstruction's significance. 
Quantitatively, it demonstrates that the overlap reconstruction substantially alleviates the blocking artifacts in the models' SR output, which can also be seen in Figure~\ref{fig:OR_BA}. 
As shown in Table~\ref{tab:main_results}, PSNRB increases significantly by an average of $\sim$0.14 dB. 
Our loss function improves the PSNR and PSNRB by $0.054$ dB and $0.039$ dB (compare row $4$,$6$). 
\begin{table*}[ht!]
    \Large
    \centering
    \caption{Comparison of proposed algorithm and RDN based on executability on low-powered devices. ``I'' - Image based prediction; ``P'' - Patch based prediction.}
    \label{tab:CRL_examination}    
    \resizebox{\textwidth}{!}{
    \begin{tabular}{|c|c|c|c|c|c|c|c|c|c|}
    \hline
        \multirow{3}{*}{Model} & \multirow{3}{*}{PSNR} & \multirow{3}{*}{PSNRB} & Total & GPU time per & CPU time per & SR image & SR image & GPU mem- & Peak Insta-\\
         & & & patches per & prediction & prediction & reconstruction & reconstruction & ory per pred- & ntaneous \\ 
         & & & image & (sec) & (sec) & time, GPU (sec) & time, CPU (sec) & iction (GB) & Power (W) \\ \hline \hline
         RDN-I & \multirow{1}{*}{$69.943$} & \multirow{1}{*}{$69.856$} &  \multirow{1}{*}{$-$} & \multirow{1}{*}{$7.321$} & \multirow{1}{*}{$15.423$} & \multirow{1}{*}{$-$} & \multirow{1}{*}{$-$} & \multirow{1}{*}{$11.943$} & \multirow{1}{*}{$210$} \\ 
         RDN-P & \multirow{1}{*}{$69.863$} & \multirow{1}{*}{$69.690$} &  \multirow{1}{*}{$361$} & \multirow{1}{*}{$0.016$} & \multirow{1}{*}{$0.064$} & \multirow{1}{*}{$6.027$} & \multirow{1}{*}{$23.374$} & \multirow{1}{*}{$6.133$} & \multirow{1}{*}{$63$} \\
         RDNLA & \multirow{1}{*}{$69.971$} & \multirow{1}{*}{$69.879$} &  \multirow{1}{*}{$1444$} & \multirow{1}{*}{$0.019$} & \multirow{1}{*}{$0.169$} & \multirow{1}{*}{$27.580$} & \multirow{1}{*}{$244.290$} & \multirow{1}{*}{$6.243$} & \multirow{1}{*}{$69$} \\\hline
    \end{tabular}
    }
\end{table*}

Table~\ref{tab:CRL_examination} shows a comparison of RDN-I, RDN-P and RDNLA based on several parameters. All the models are evaluated on RTX-5000 NVIDIA GPU. The effect of the input size can be understood by comparing  ``GPU memory per prediction'' and ``peak instantaneous power'' of RDN-I and RDN-P. The memory occupied and power consumed decrease by 49\% and 70\%, respectively. However, the model's SR performance (PSNR, PSNRB) decreases, which is not preferred. Our proposed SR algorithm (RDNLA) provides better SR performance than the other models, consumes  48\% lesser memory and 67\% lesser power than the baseline, RDN-I. 

The power available for satellite subsystems depends on various factors such as the type of satellite mission, number of subsystems and type of solar panels. 
For example, a solar panel mounted on Cubesat~\cite{claricoats2018design} has 7~W rating, commercial LEO satellites~\cite{kharsanky2017power} consume 40~W, whereas LRO spacecraft power is 685~W and Selene's main orbiter's power is 3.5~kW~\cite{haruyama2008global}. Our proposed SR algorithm's instantaneous power consumption can be increased/decreased by increasing/decreasing the patch size, making it executable even on satellites with lower power availability. Though our algorithm provides enhanced results while consuming less power and memory, the ``SR image reconstruction time'' is significantly increased due to an increase in patches, as shown in Table~\ref{tab:CRL_examination}. Future research could address this problem by proposing real-time SR algorithms that provide enhanced SR results and are executable on low-powered devices. 

\section{Ablation Investigation}
\textbf{Performance on natural images.} We also train our network on the DIV2K dataset~\cite{Timofte_2017_CVPR_Workshops} to observe our algorithm's performance on natural images.
We follow the exact training settings as mentioned by~\cite{Zhang_2020_TPAMI}. 
We trained our model for a total of $800$ epochs and utilize $\text{P}=16$, $\text{D}=8$, and $\text{G}=64$ to report our results. We also employ a self-ensemble strategy~\cite{Zhang_2020_TPAMI} to boost the performance of RDNLA (RDNLA+).
Table~\ref{tab:set5_results} shows the quantitative comparisons for $\text{x}2, ~\text{x}3, ~\text{and} ~\text{x}4$ SR on the Set5~\cite{set5}, Set14~\cite{set14}, and B100~\cite{b100} datasets. It can be observed that our proposed algo. outperforms all low-powered device SR models significantly but provides slightly lower performance than standard SR models. This may be because the proposed loss function lays little emphasis on the image's boundary regions.
\begin{table*}[ht!]
    \centering
    \caption{Quantitative results (PSNR/SSIM) on natural image SR benchmark datasets. Best and second best results are \textbf{highlighted} and \underline{underlined}.}
    \label{tab:set5_results}
    \resizebox{\textwidth}{!}
    {
    \begin{tabular}{|c|c|c|c|c|c|c|c|c|c|}
    \hline
       \multirow{2}{*}{Method} & \multicolumn{3}{c|}{Set5} & \multicolumn{3}{c|}{Set14} & \multicolumn{3}{c|}{B100} \\ \cline{2-10}
       & $\times2$ & $\times3$ & $\times4$ & $\times2$ & $\times3$ & $\times4$ & $\times2$ & $\times3$ & $\times4$ \\ \hline \hline
        \multicolumn{10}{|c|}{\textbf{Standard SR}}\\\hline
        RDN~\cite{Zhang_2020_TPAMI} & $38.24/0.9614$ & $34.71/0.9296$ & $32.47/0.8990$ & $34.01/0.9212$ & $30.57/0.8468$ & $28.81/0.7871$ & $32.34/0.9017$ & $29.26/0.8093$ & $27.72/0.7419$\\
        RCAN~\cite{RCAN} & $\underline{38.27}/\underline{0.9614}$ & $\underline{34.74}/\underline{0.9299}$ & $\underline{32.63}/\underline{0.9002}$ & $\textbf{34.12}/\textbf{0.9216}$ & $\textbf{30.65}/\textbf{0.8482}$ & $\underline{28.87}/\textbf{0.7889}$ & $\underline{32.41}/\underline{0.9027}$ & $\underline{29.32}/\underline{0.8111}$ & $\underline{27.77}/\textbf{0.7436}$\\
        SRFBN~\cite{SRFBN} & $38.11/0.9609$ & $34.70/0.9292$ & $32.47/0.8983$ & $33.82/0.9196$ & $30.51/0.8461$ & $28.81/0.7868$ & $32.29/0.9010$ & $29.24/0.8084$ & $27.72/0.7409$\\
        RNAN~\cite{RNAN} & $38.17/0.9611$ & - & $32.49/0.8982$ & $33.87/0.9207$ & - & $28.83/0.7878$ & $32.32/0.9014$ & - & $27.72/0.7421$ \\       
        OISR~\cite{OISR} & $38.21/0.9612$ & $34.72/0.9297$ & $32.53/0.8992$ & $33.94/0.9206$ & $30.57/0.8470$ & $28.86/0.7878$ &  $32.36/0.9019$ & $29.29/0.8103$ & $27.75/\underline{0.7428}$ \\
        SAN~\cite{SAN} & $\textbf{38.31}/\textbf{0.9620}$ & $\textbf{34.75}/\textbf{0.9300}$ & $\textbf{32.64}/\textbf{0.9003}$ & $\underline{34.07}/\underline{0.9213}$ & $\underline{30.59}/\underline{0.8476}$ & $\textbf{28.92}/\underline{0.7888}$ & $\textbf{32.42}/\textbf{0.9028}$ & $\textbf{29.33}/\textbf{0.8112}$ & $\textbf{27.78}/\textbf{0.7436}$\\ \hline \hline
        \multicolumn{10}{|c|}{\textbf{Low Powered Device SR}}\\\hline
        MobiSR~\cite{lee2019mobisr} & - & - & $31.73/0.8873$ & - & - & $28.24/0.7729$ & - & - & $27.33/0.7283$ \\
        Kim et al.~\cite{kim2018real} & $36.66/0.9548$ & - & - & $32.52/0.9073$ & - & - & $31.32/0.8880$ & - & - \\
        SRNPU~\cite{lee2020srnpu} & $37.06/0.9565$ & $32.62/0.9099$ & $31.47/0.8893$ & $33.59/0.9258$ & $29.74/0.8399$ & $28.92/0.8013$ & $30.41/0.8578$ & $27.37/0.7534$ & $26.86/0.7079$ \\        
        SplitSR~\cite{liu2021splitsr} & - & - & $31.76/0.8982$ & - & - & $28.29/0.7916$ & - & - & $27.39/0.7491$ \\
        MFAGAN~\cite{cheng2021mfagan} & - & - & $30.16/-$ & - & - & $26.69/-$ & - & - & $25.33/-$ \\        
        \hline
        RDNLA & $38.12/0.9611$ & $34.53/0.9286$ & $32.25/0.8963$ & $33.76/0.9196$ & $30.42/0.8451$ & $28.68/0.7840$ & $32.27/0.9010$& $29.17/0.8080$ & $27.63/0.7384$\\
        RDNLA+ & $38.22/0.9615$ & $34.68/0.9298$ & $32.46/0.8987$ & $33.86/0.9205$ & $30.56/0.8469$ & $28.82/0.7868$ & $32.33/0.9017$ & $29.26/0.8095$ & $27.72/0.7404$\\ \hline
    \end{tabular}
    }
\end{table*}

\begin{figure*}[!htb]
\resizebox{0.78\textwidth}{!}{
\begin{minipage}[!ht]{.85\textwidth}
\resizebox{0.9\textwidth}{!}{
\includegraphics[height = \textwidth]{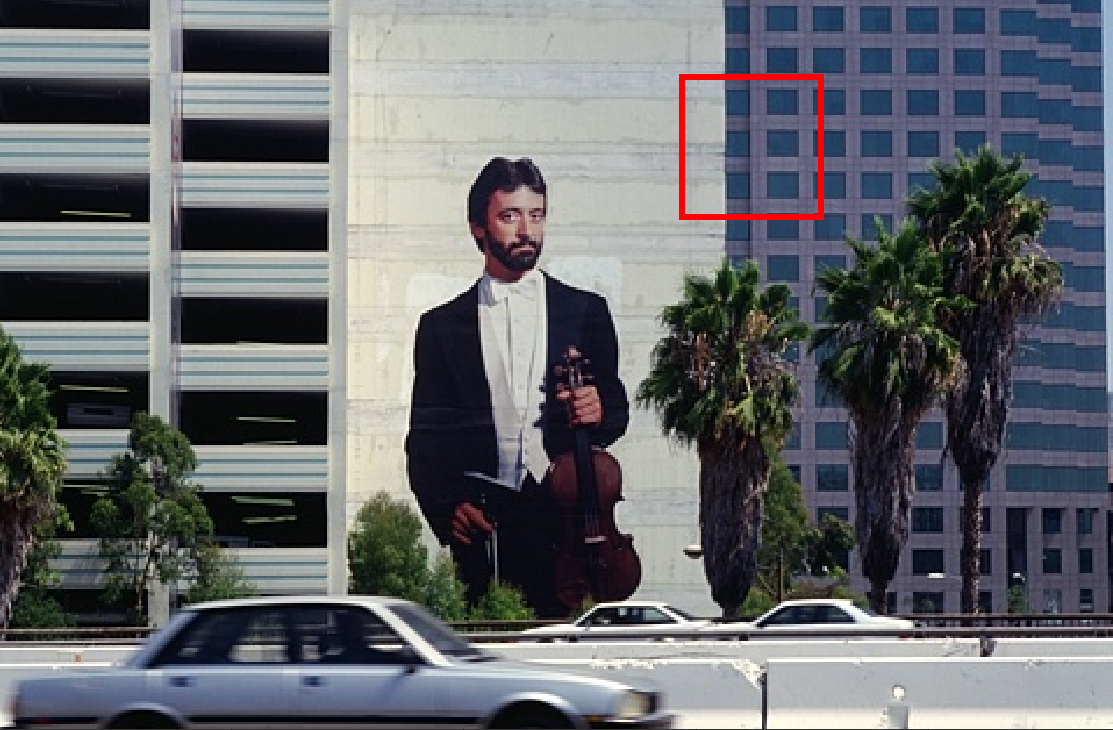}
}
\end{minipage}
\begin{minipage}[t]{.7\textwidth}
\begin{tabular}{ccccc}
\centering
\fcolorbox{red}{yellow}{\includegraphics[height=0.3\textwidth]{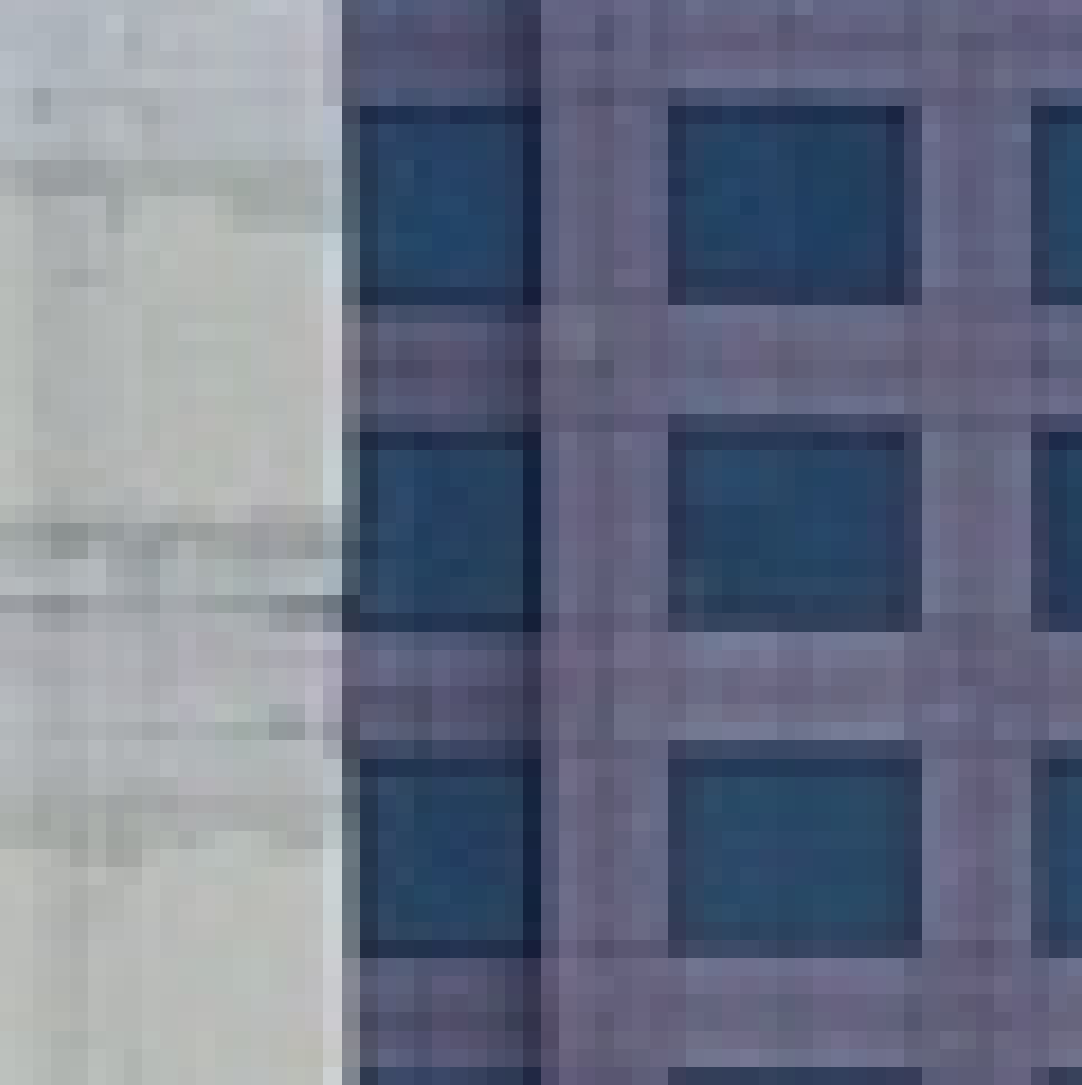}} &
\fcolorbox{red}{yellow}{\includegraphics[width=0.3\textwidth]{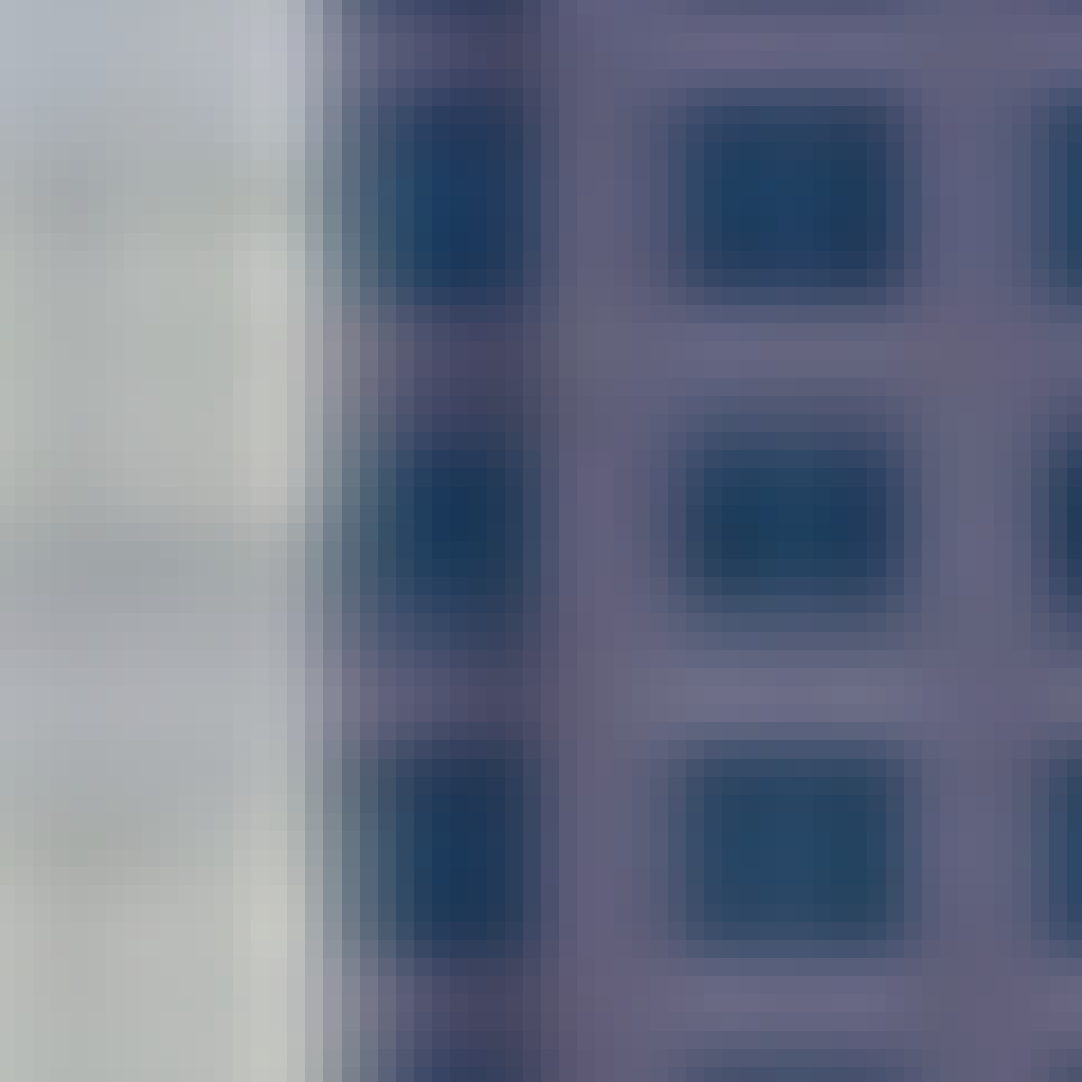}} &
\fcolorbox{red}{yellow}{\includegraphics[width=0.3\textwidth]{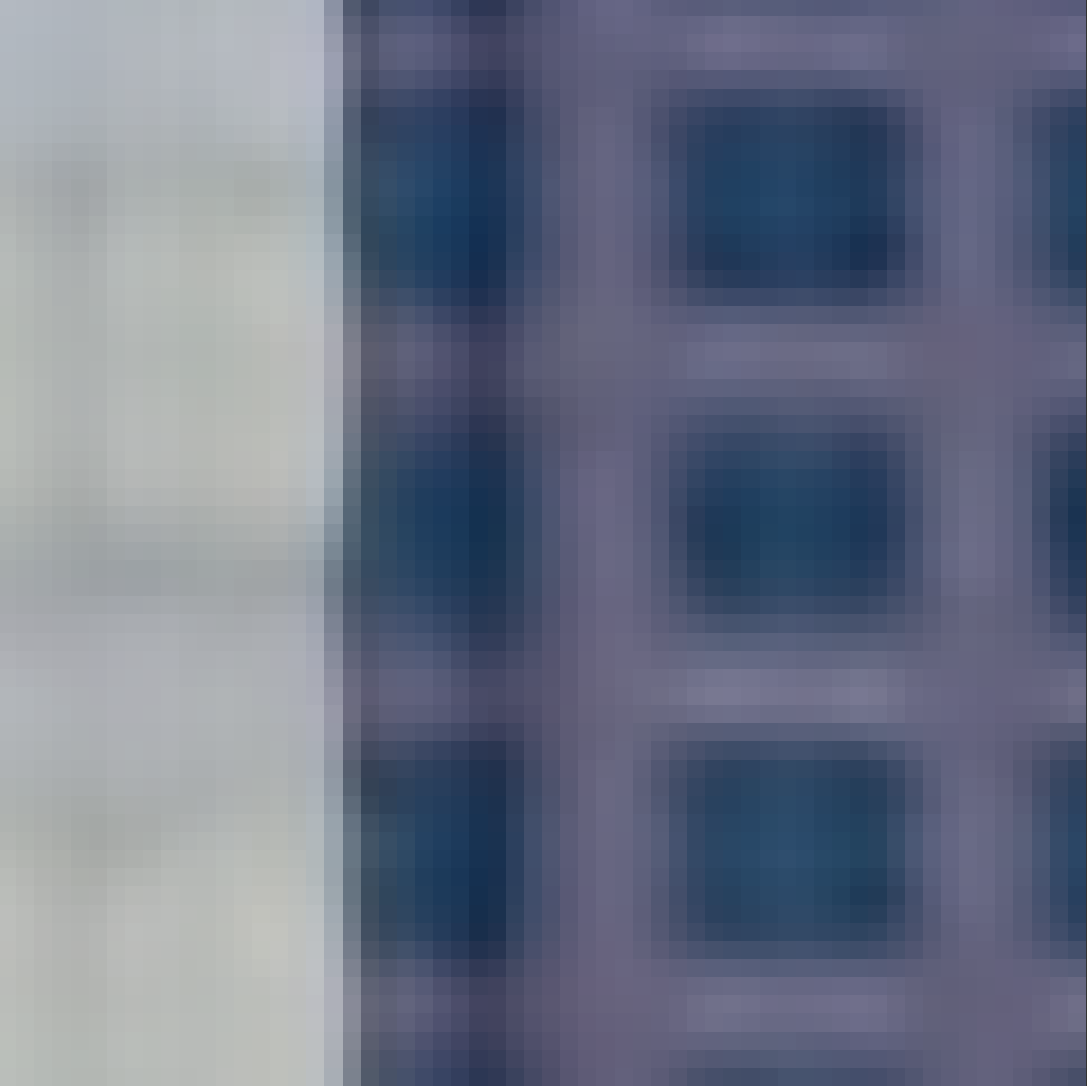}}  &
\fcolorbox{red}{yellow}{\includegraphics[width=0.3\textwidth]{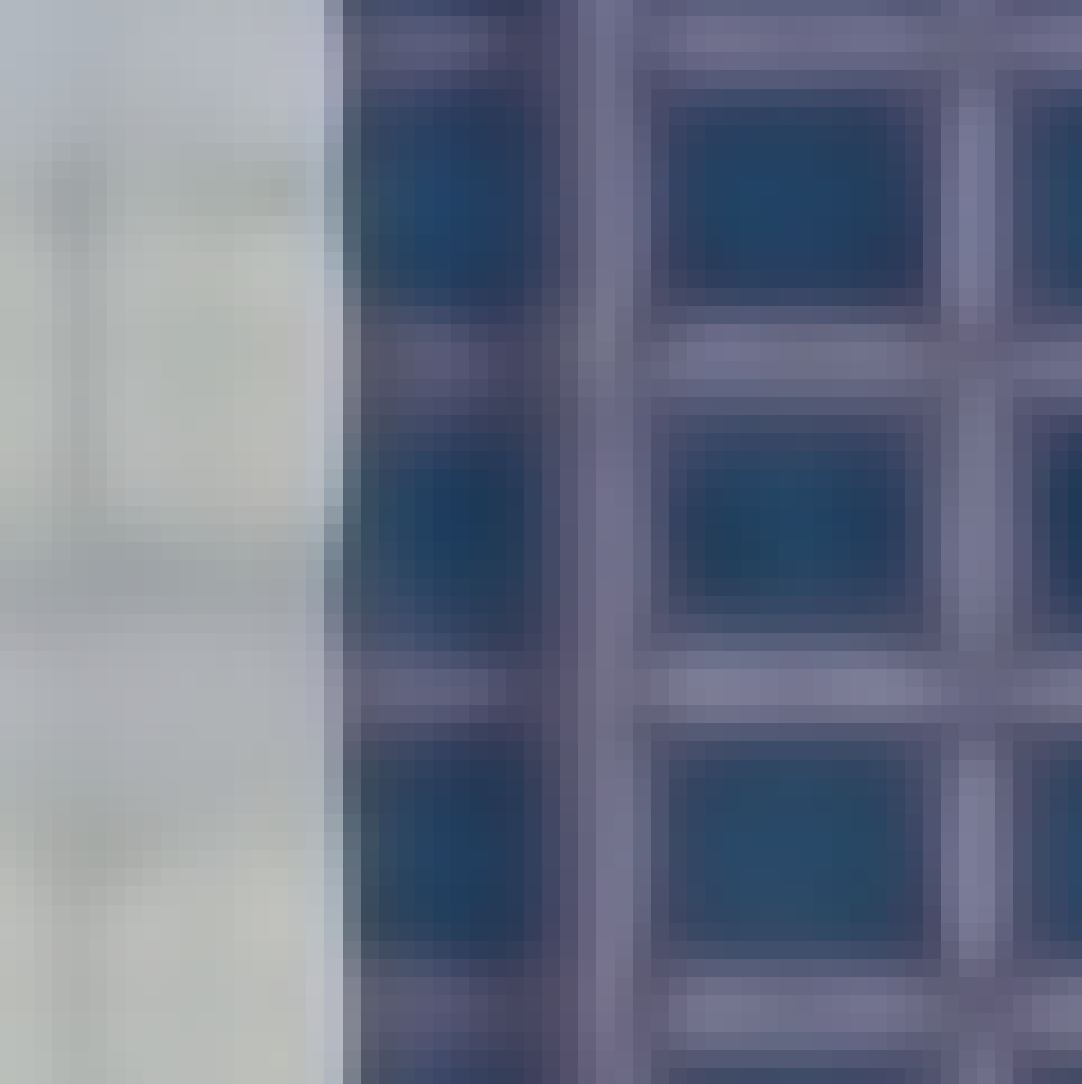}} &
\fcolorbox{red}{yellow}{\includegraphics[width=0.3\textwidth]{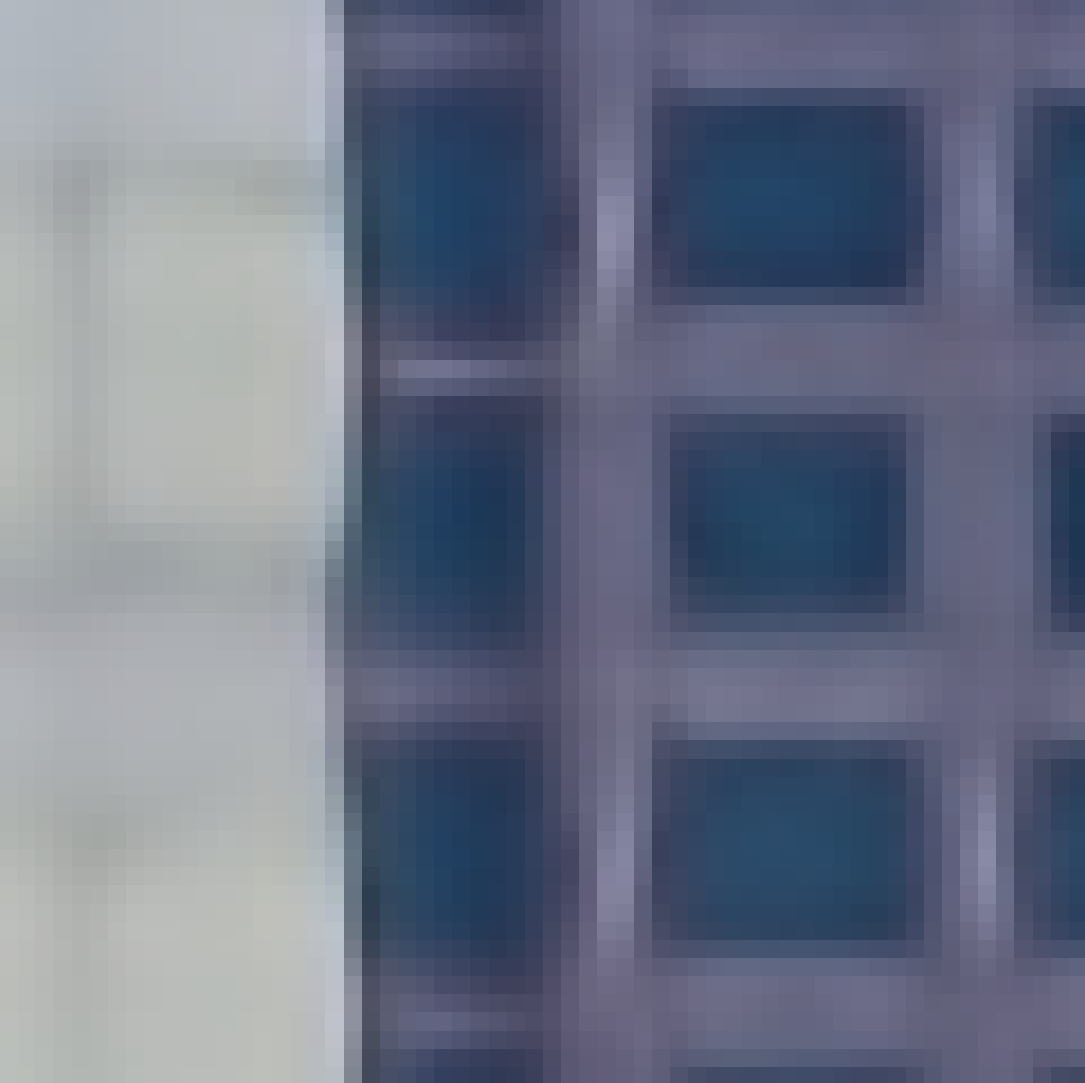}} 
\\
HR &
Bicubic &
SRCNN~\cite{dong2015image}  &
DRCN~\cite{Kim_2016_CVPR1}  &
VDSR~\cite{Kim_2016_CVPR}
\\
&&&&
\\
\fcolorbox{red}{yellow}{\includegraphics[width=0.3\textwidth]{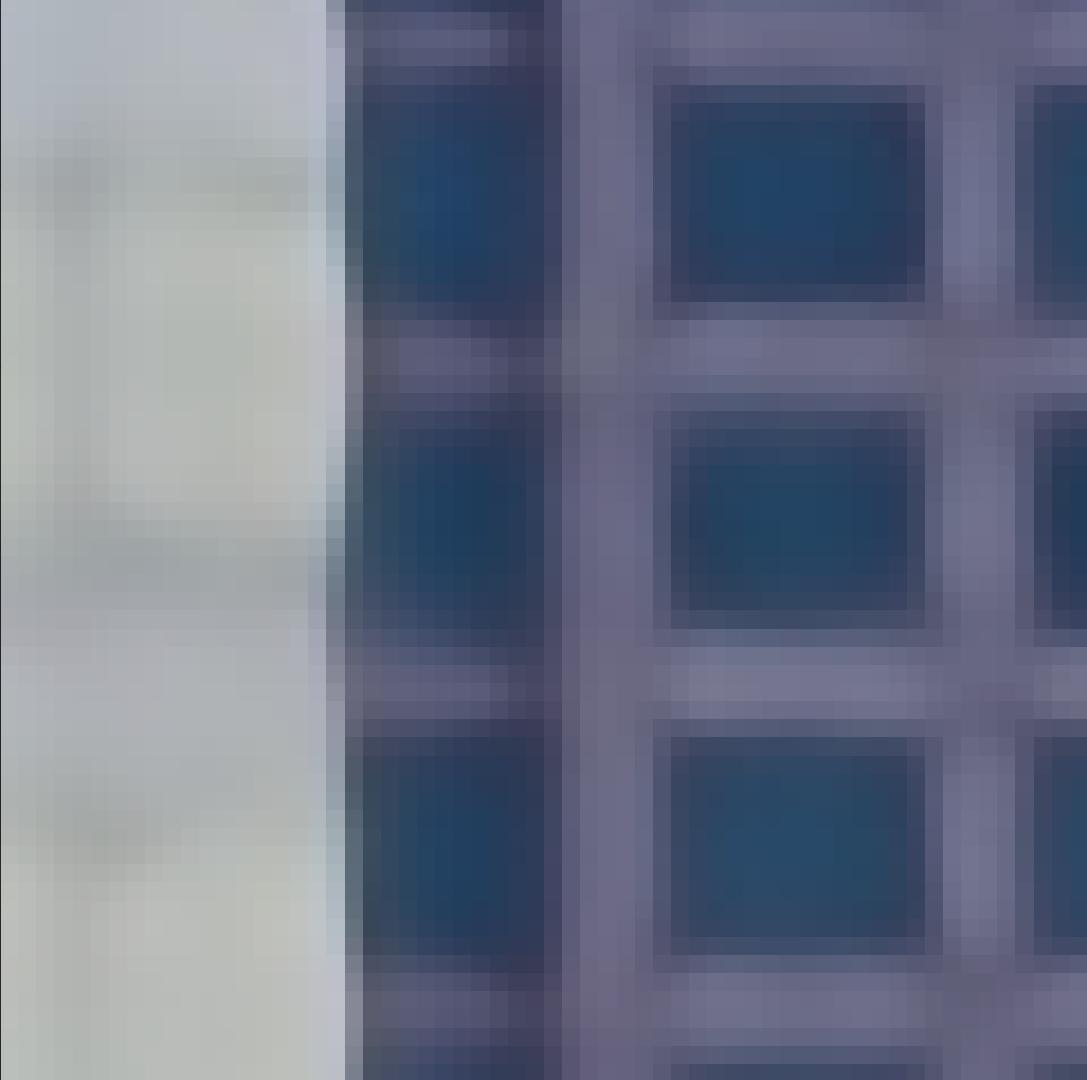}} &
\fcolorbox{red}{yellow}{\includegraphics[width=0.3\textwidth]{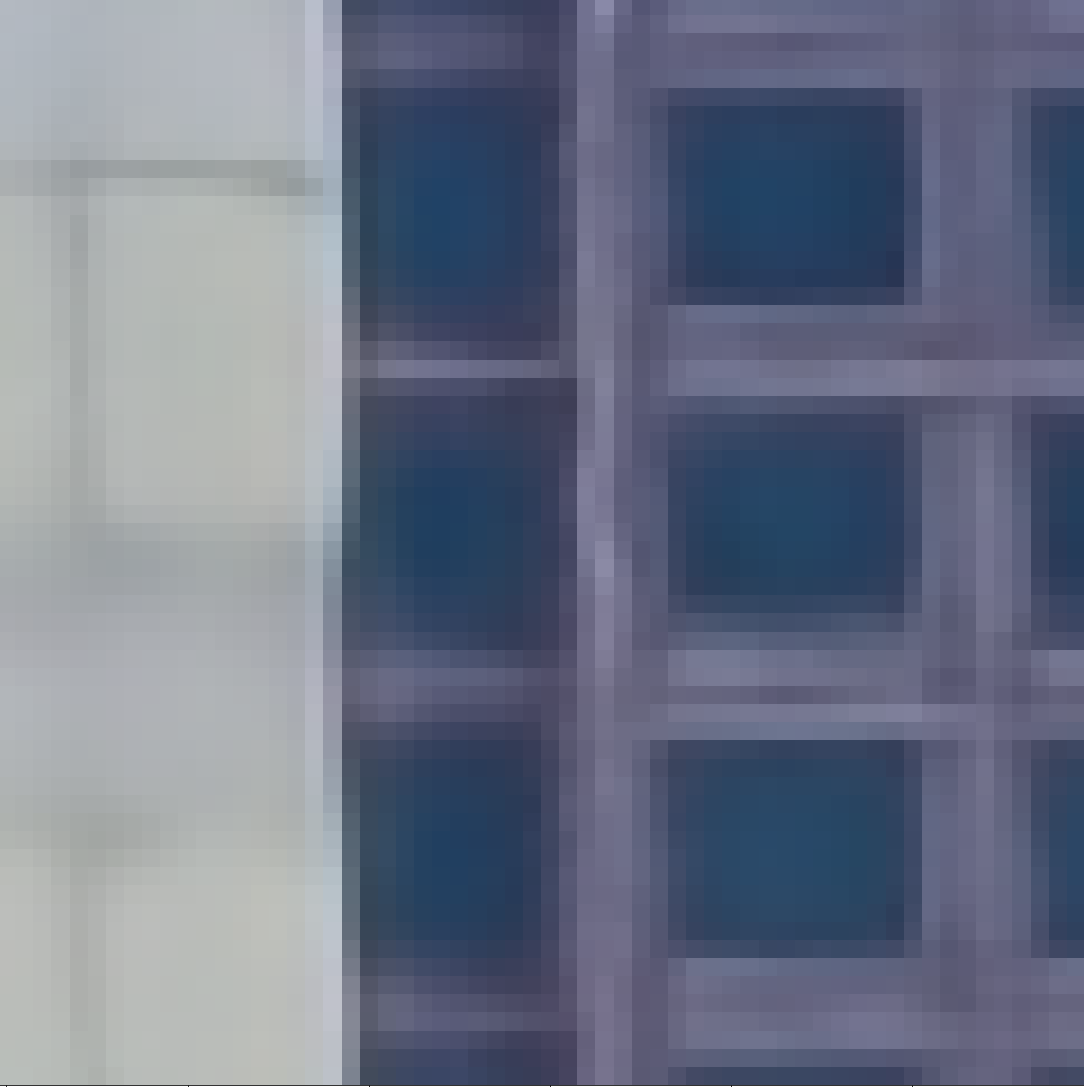}} &
\fcolorbox{red}{yellow}{\includegraphics[width=0.3\textwidth]{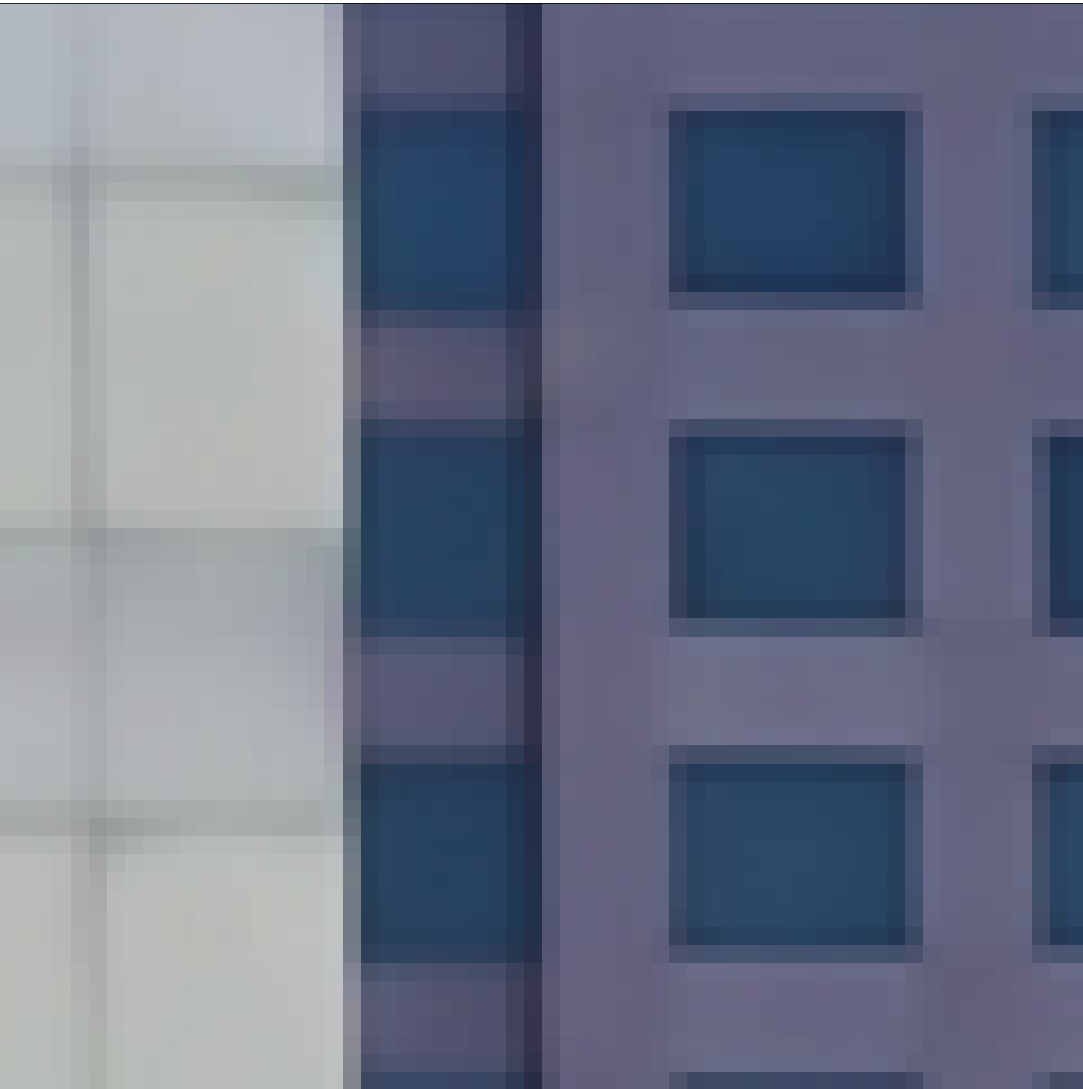}}  & 
\fcolorbox{red}{yellow}{\includegraphics[width=0.3\textwidth]{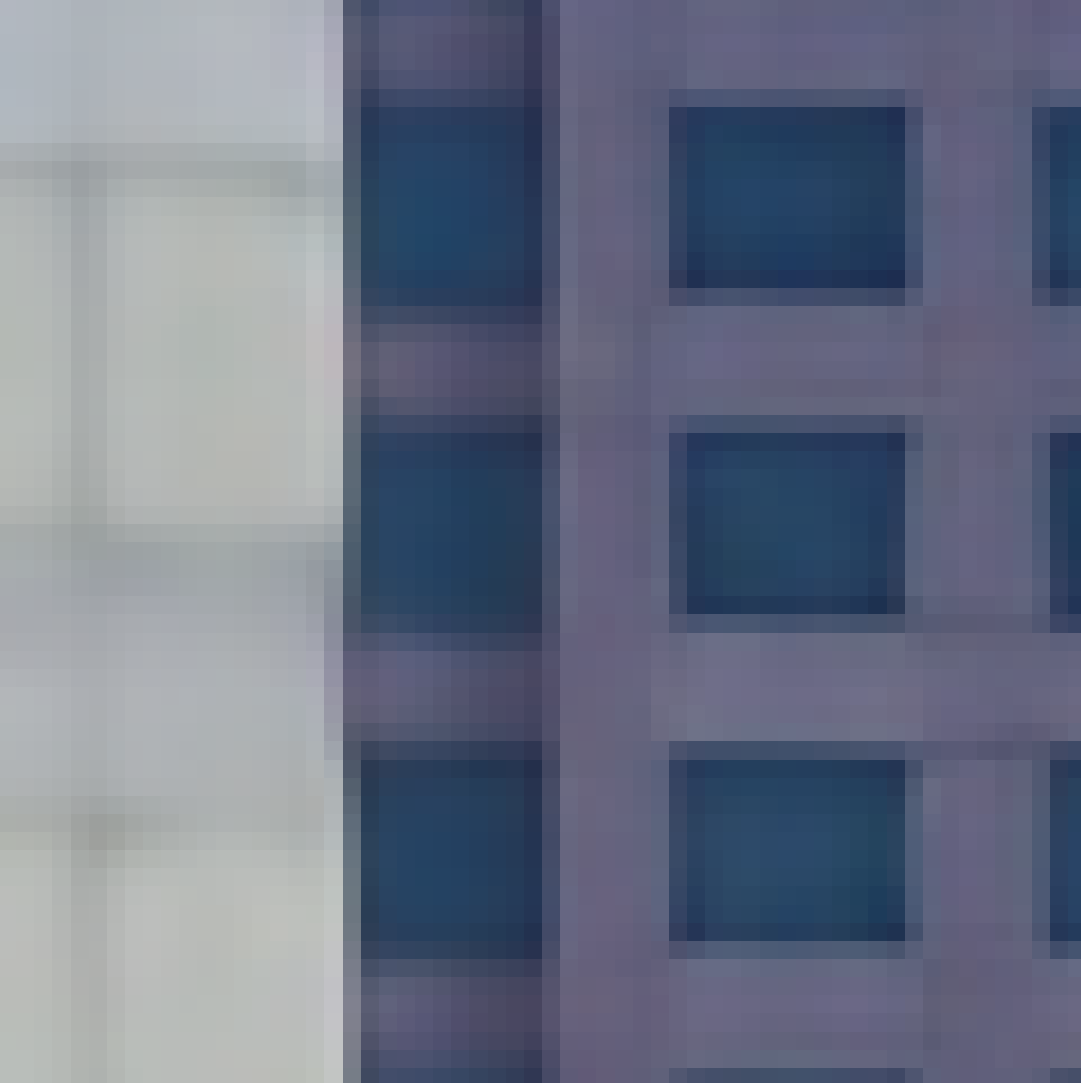}}  &
\fcolorbox{red}{yellow}{\includegraphics[width=0.3\textwidth]{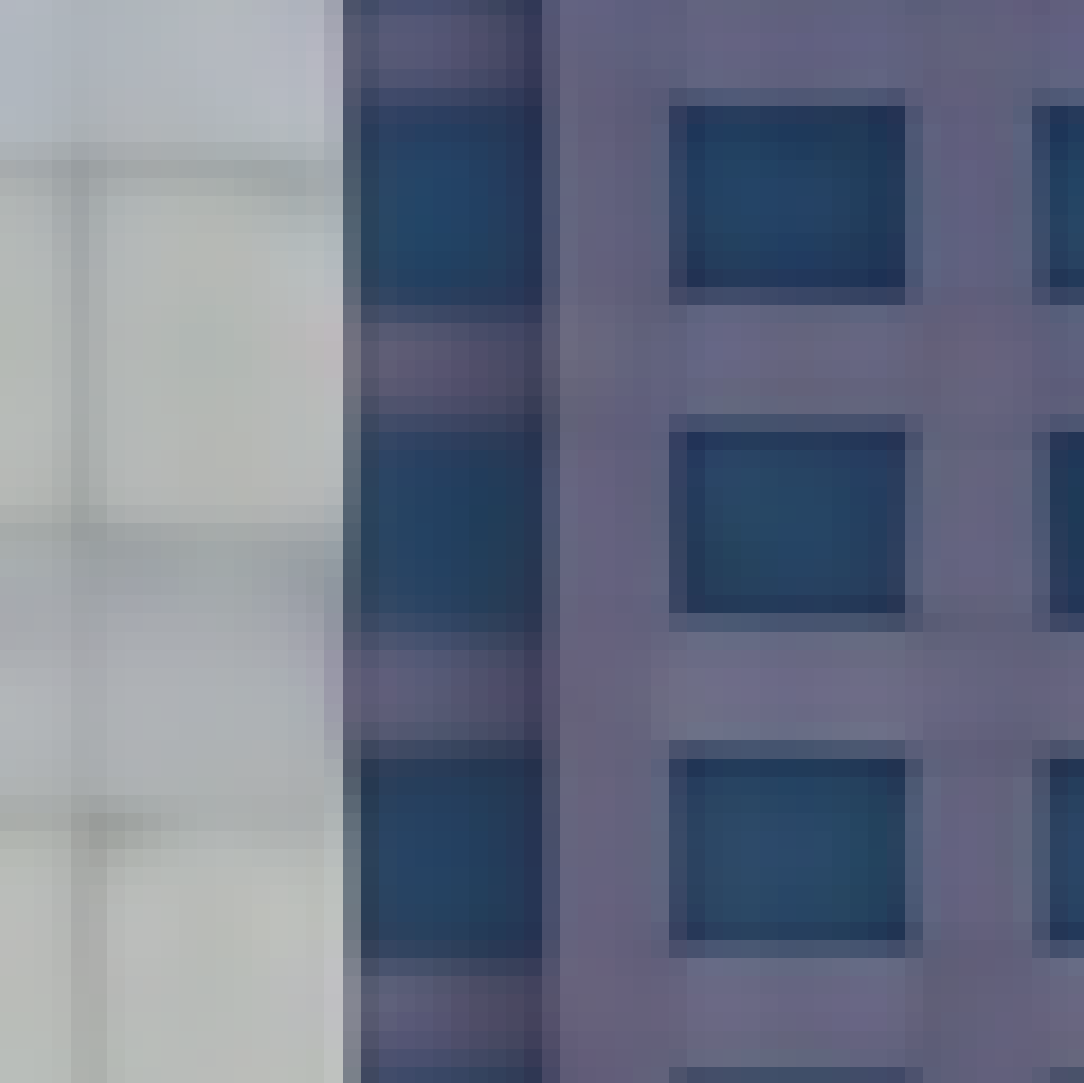}}  
\\
LapSRN~\cite{lai2017deep} &
DRRN~\cite{Tai_2017_CVPR} &
D-DBPN~\cite{Haris_2018_CVPR}  &
RDNLA (ours) &
RDNLA+ (ours)
\end{tabular}
\end{minipage}
}
\caption{Visual results ($4\times$) with model trained on DIV2K~\cite{Timofte_2017_CVPR_Workshops}. The results are presented on image ``$119082$'' from B100~\cite{b100} dataset.}
\label{fig:119082}
\end{figure*}

\begin{figure*}[ht!]
\resizebox{0.8\textwidth}{!}
{
\begin{minipage}[!ht]{.36\textwidth}
{
\includegraphics[height = \textwidth]{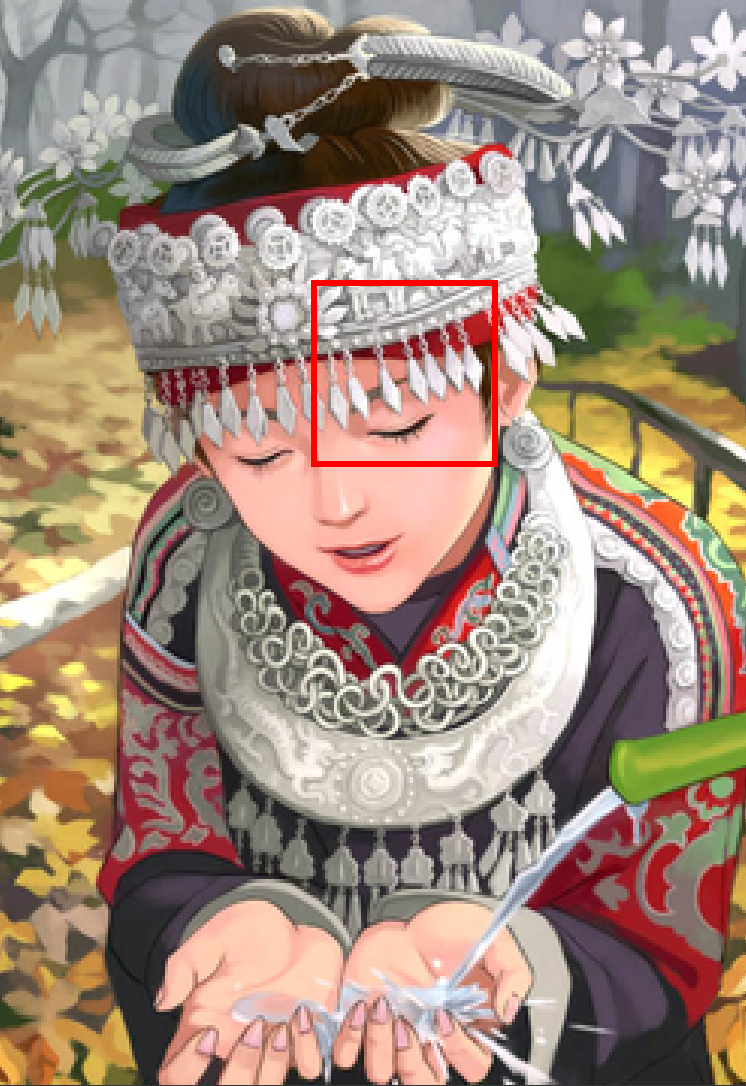}
}
\end{minipage}
\begin{minipage}[t]{.69\textwidth}
\begin{tabular}{ccccc}
\centering
\fcolorbox{red}{yellow}{\includegraphics[height=0.25\textwidth]{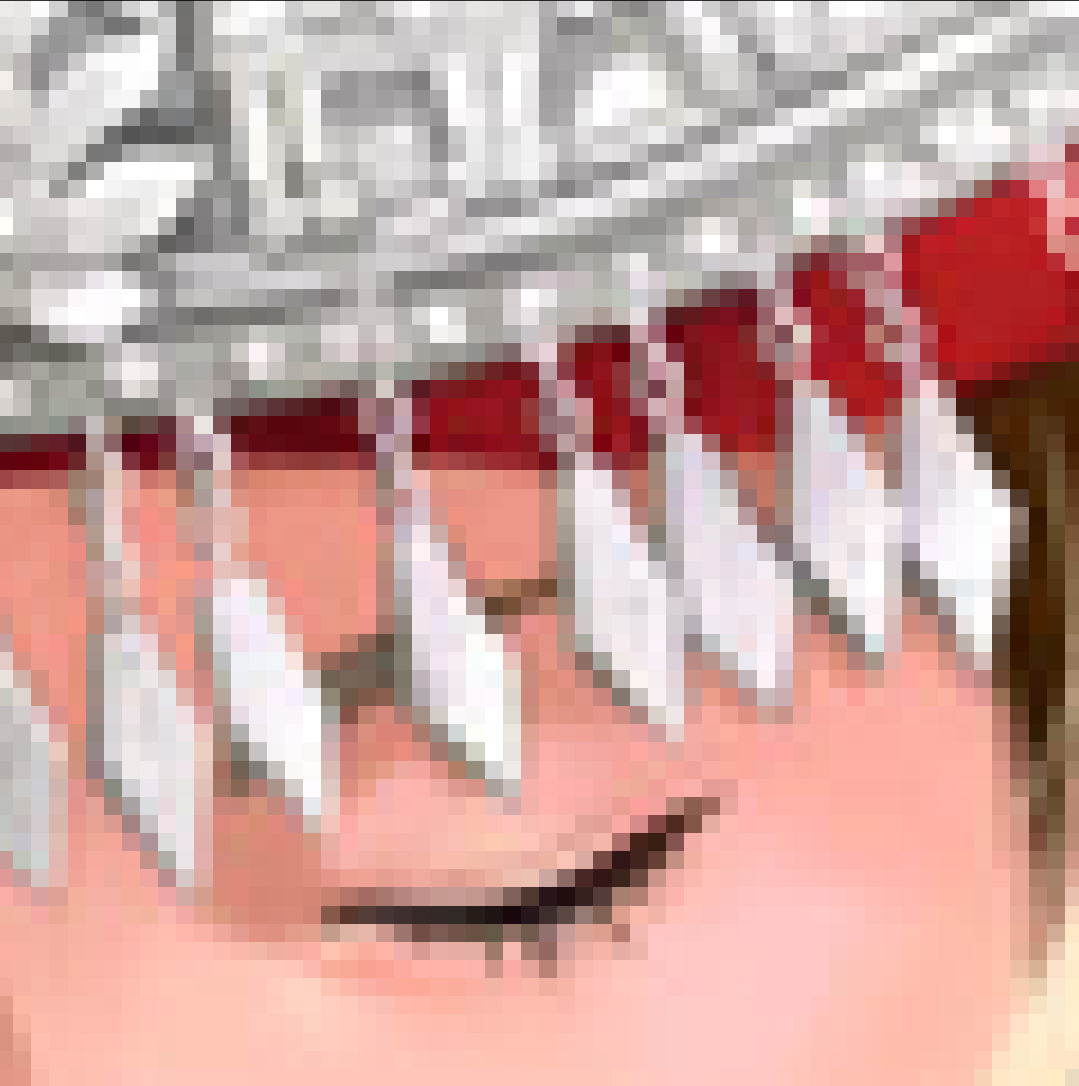}} &
\fcolorbox{red}{yellow}{\includegraphics[width=0.25\textwidth]{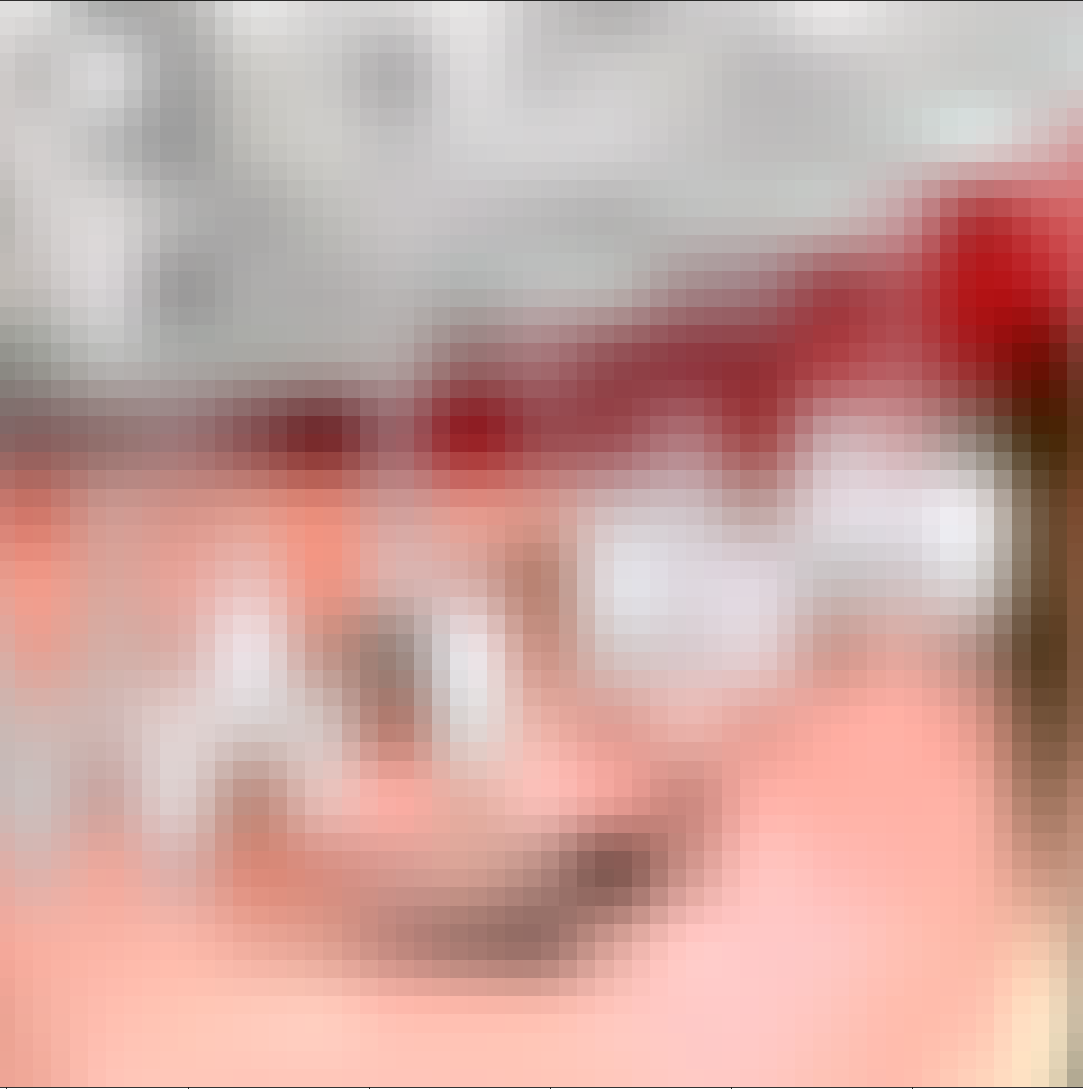}} &
\fcolorbox{red}{yellow}{\includegraphics[width=0.25\textwidth]{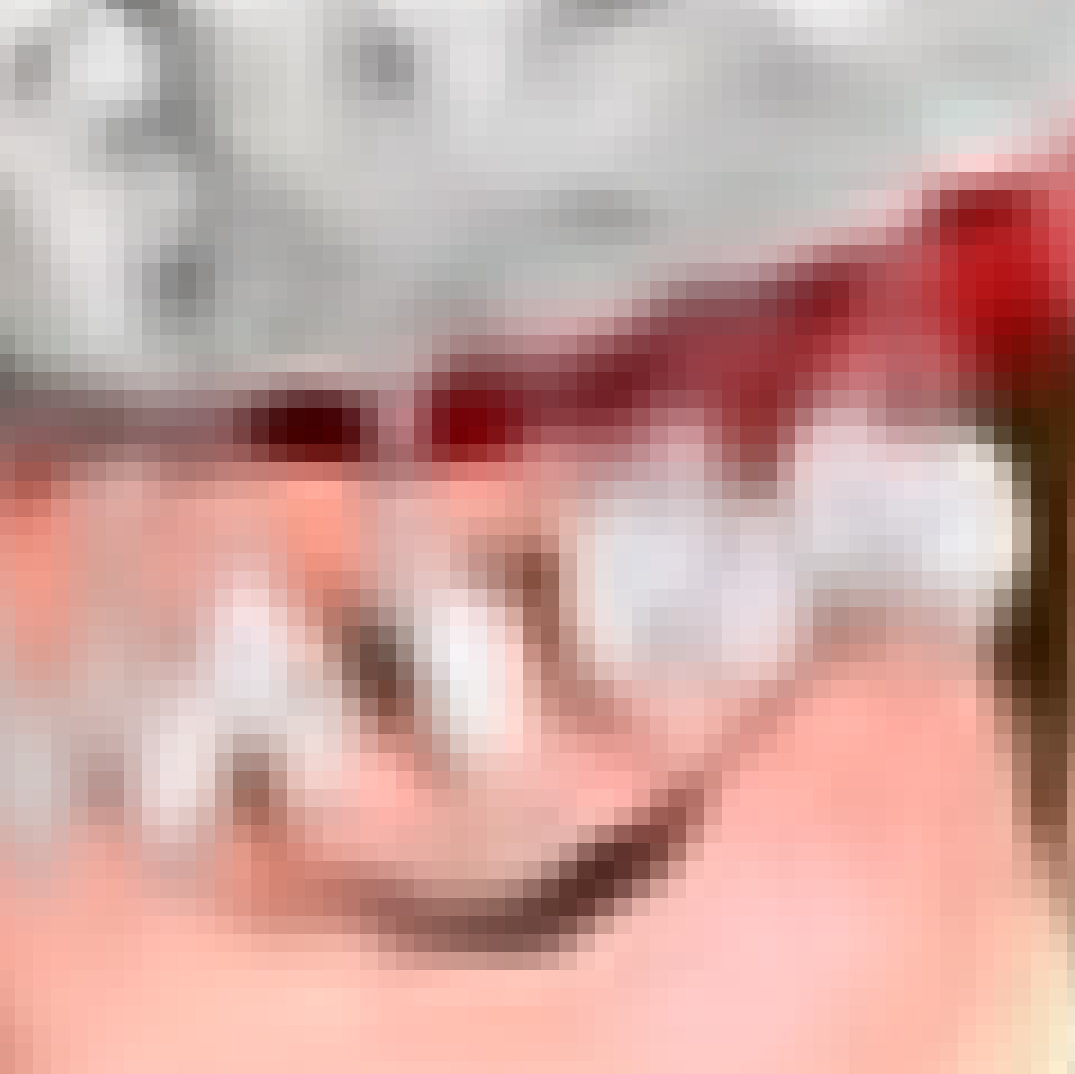}}  &
\fcolorbox{red}{yellow}{\includegraphics[width=0.25\textwidth]{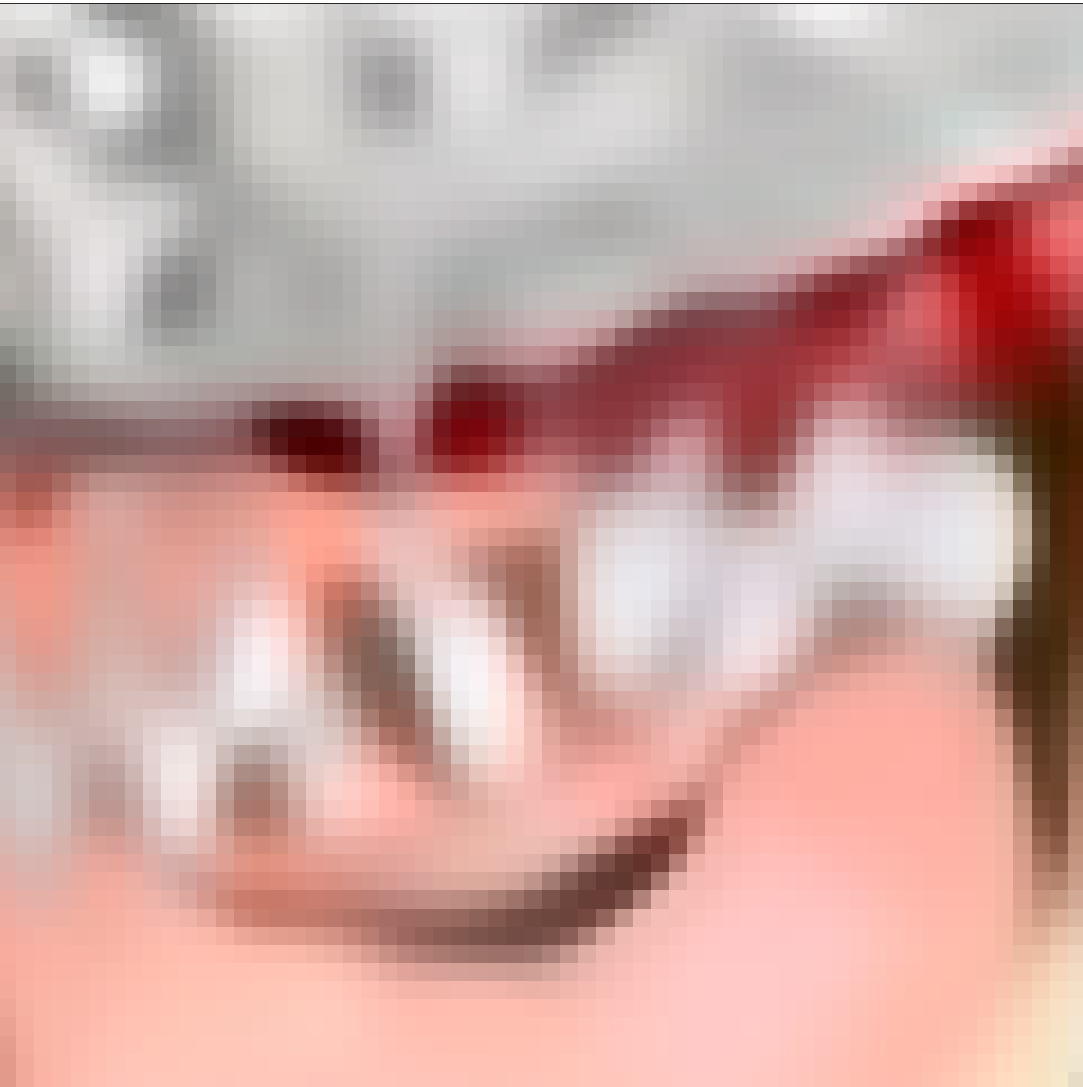}} &
\fcolorbox{red}{yellow}{\includegraphics[width=0.25\textwidth]{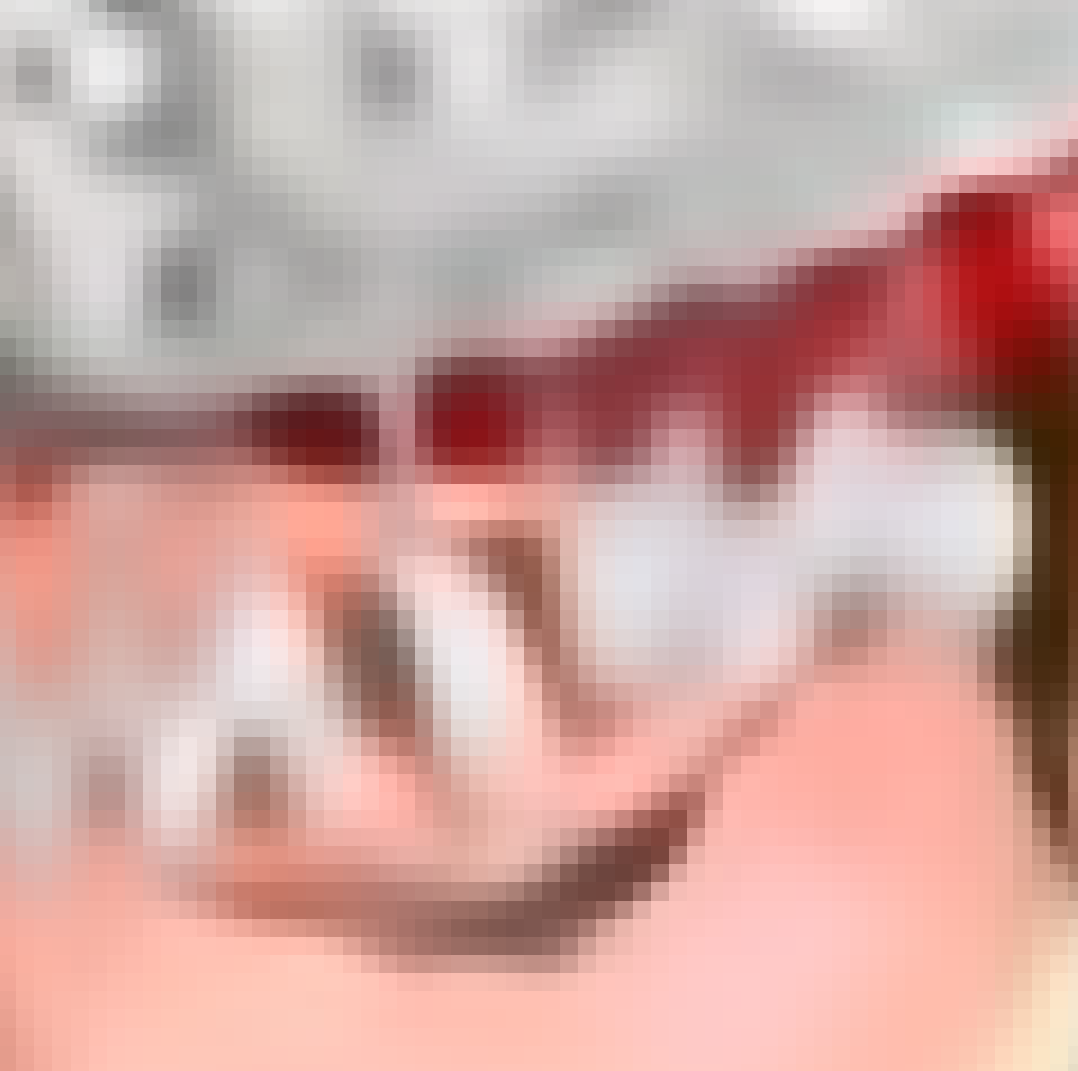}} 
\\
HR &
Bicubic &
SRCNN~\cite{dong2015image}  &
DRCN~\cite{Kim_2016_CVPR1}  &
VDSR~\cite{Kim_2016_CVPR}
\\
&&&&
\\
\fcolorbox{red}{yellow}{\includegraphics[width=0.25\textwidth]{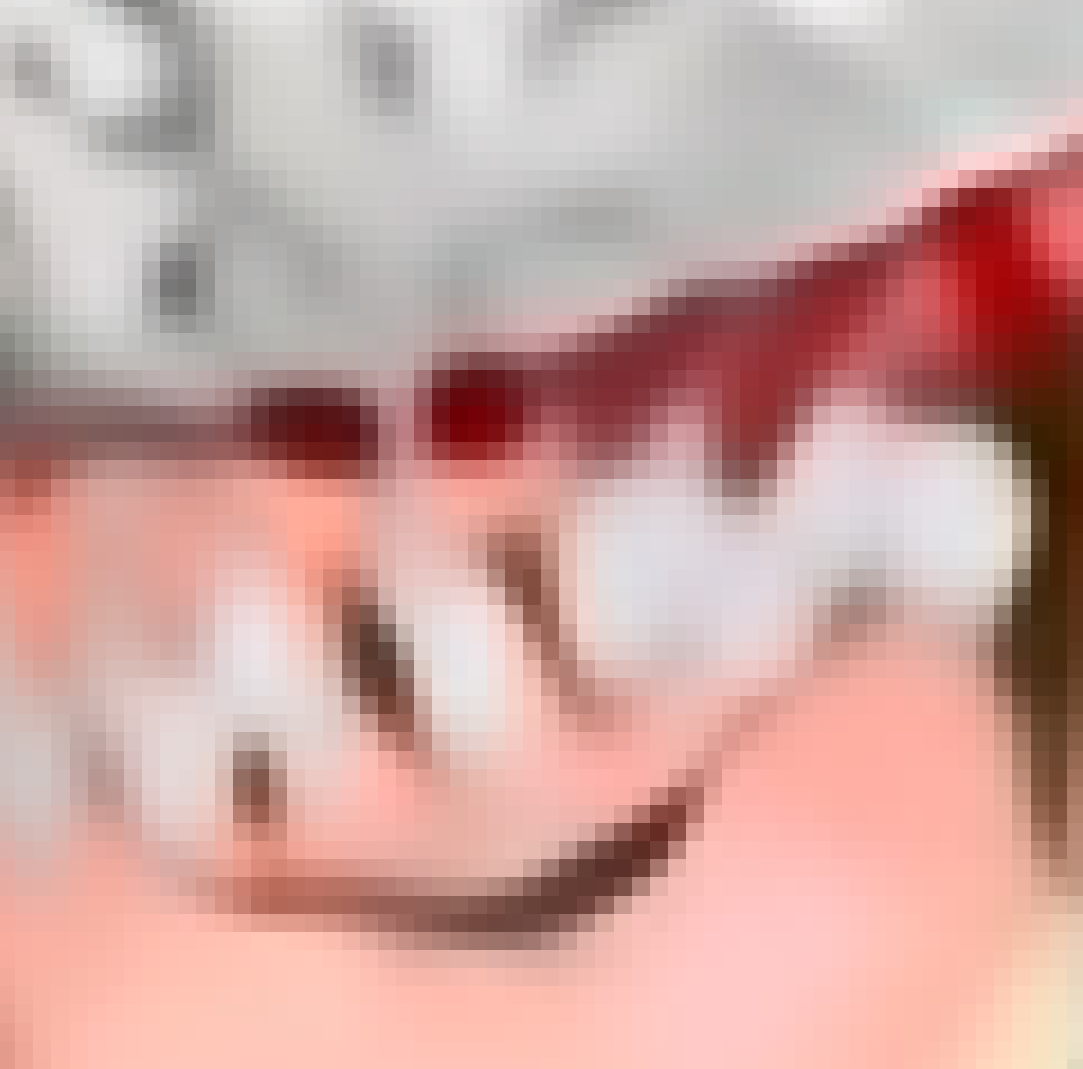}} &
\fcolorbox{red}{yellow}{\includegraphics[width=0.25\textwidth]{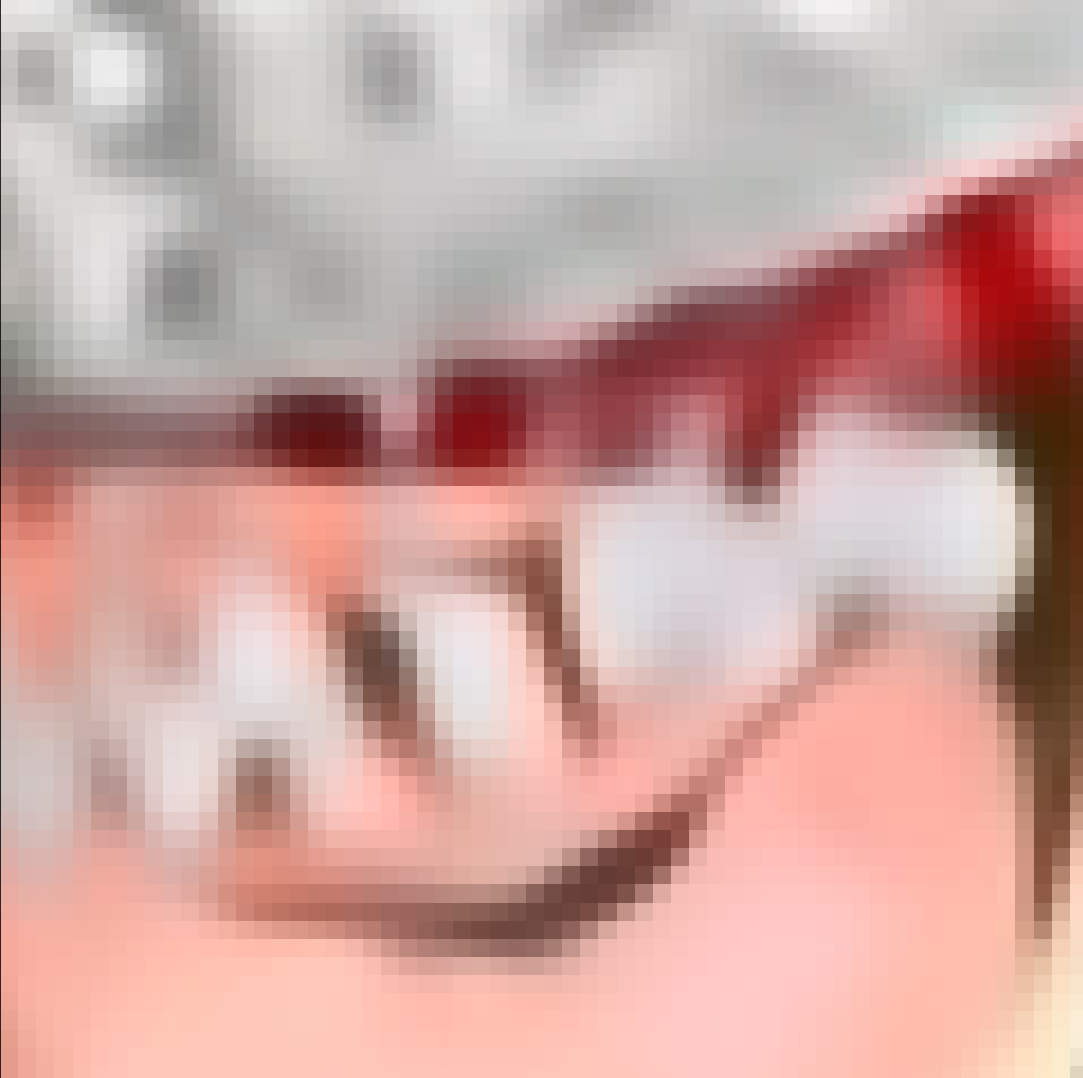}} &
\fcolorbox{red}{yellow}{\includegraphics[width=0.25\textwidth]{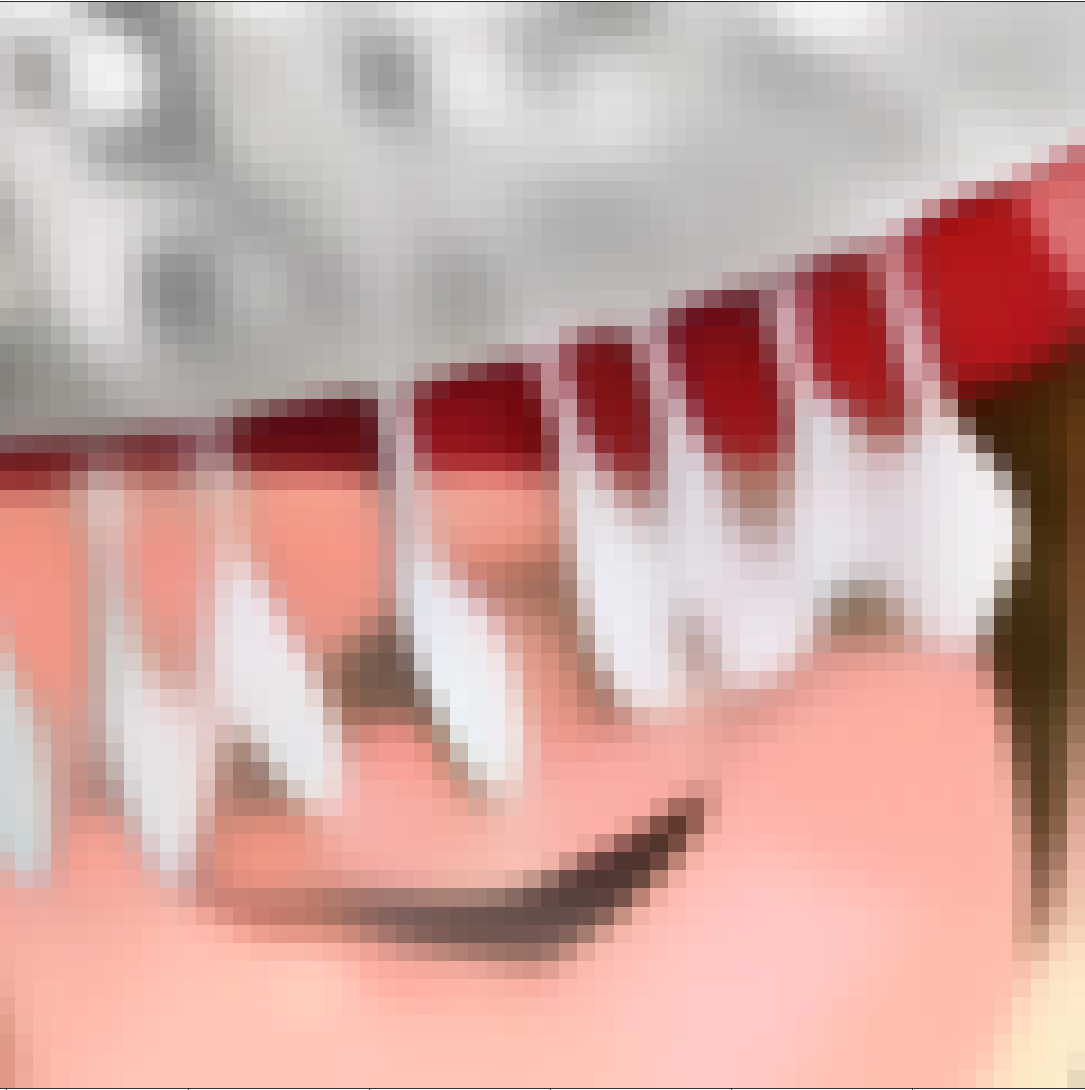}}  & 
\fcolorbox{red}{yellow}{\includegraphics[width=0.25\textwidth]{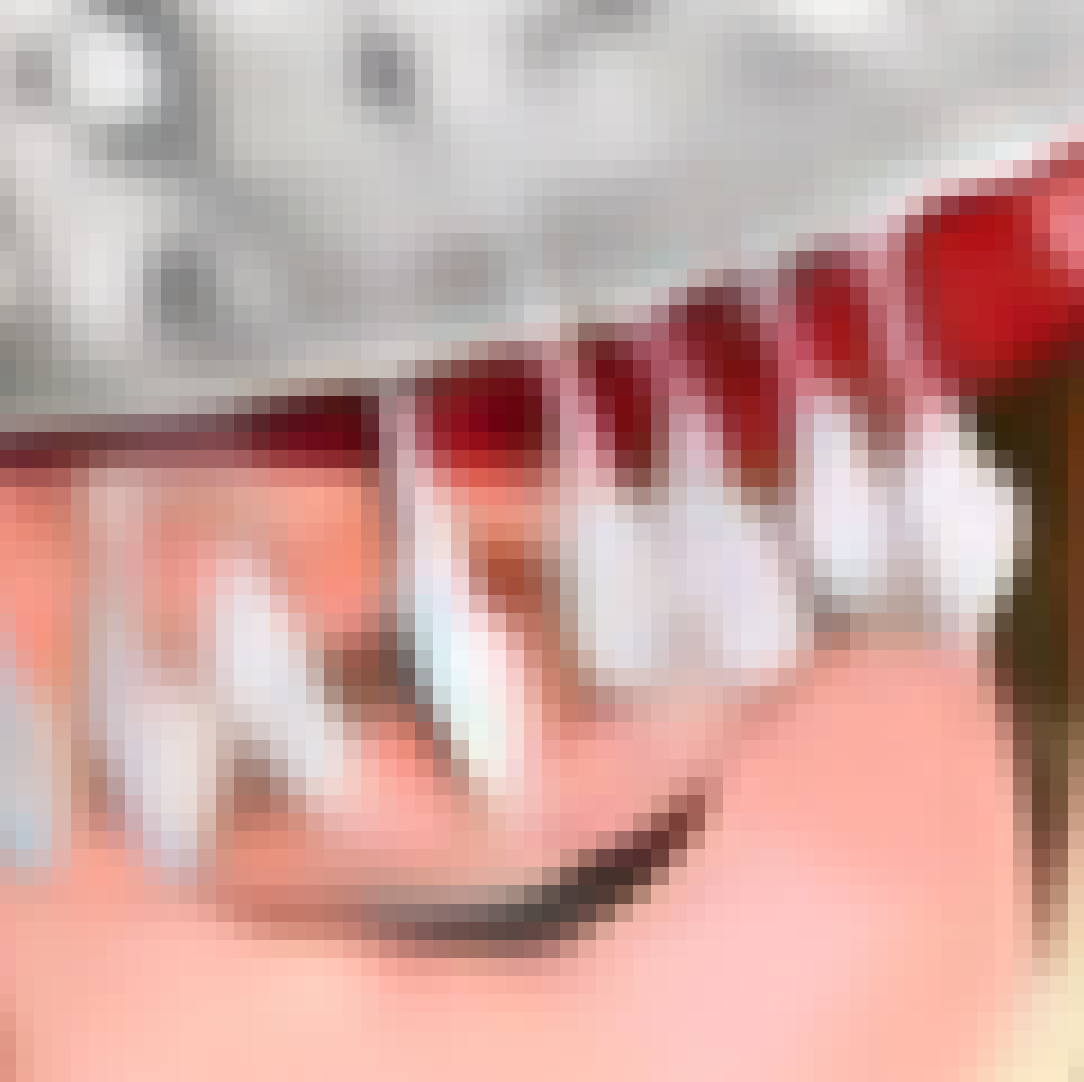}}  &
\fcolorbox{red}{yellow}{\includegraphics[width=0.25\textwidth]{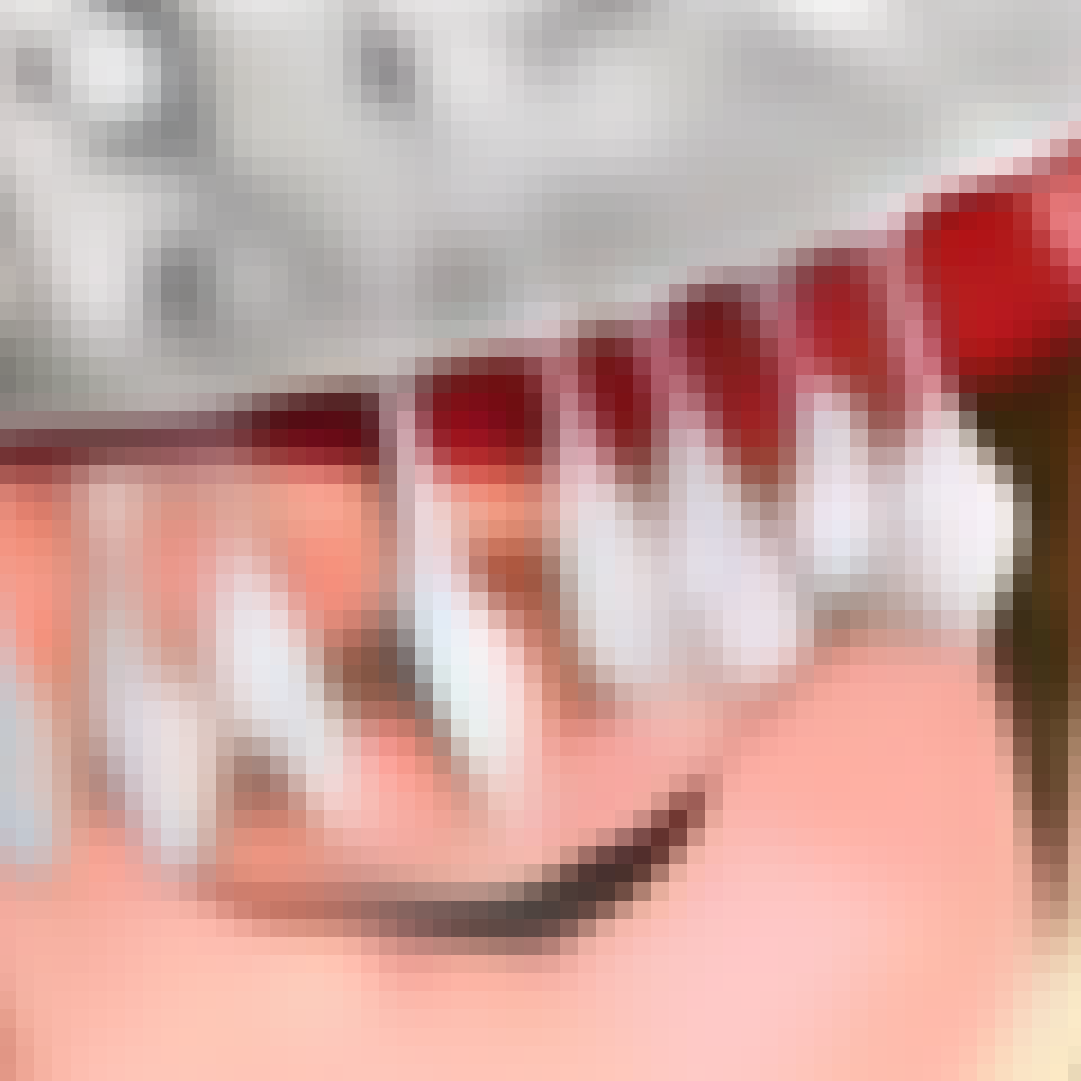}}  
\\
LapSRN~\cite{lai2017deep} &
DRRN~\cite{Tai_2017_CVPR} &
D-DBPN~\cite{Haris_2018_CVPR}  &
RDNLA (ours) &
RDNLA+ (ours)
\end{tabular}
\end{minipage}
}
\caption{Visual results ($4\times$) with model trained on DIV2K~\cite{Timofte_2017_CVPR_Workshops}. The results are presented on image ``$005$'' from Set14~\cite{set14} dataset.}
\label{fig:set14_comic}
\end{figure*}

\begin{figure*}[!htb]
\resizebox{0.8\textwidth}{!}
{
\begin{minipage}[!ht]{.36\textwidth}
\resizebox{0.9\textwidth}{!}{
\includegraphics[height = \textwidth]{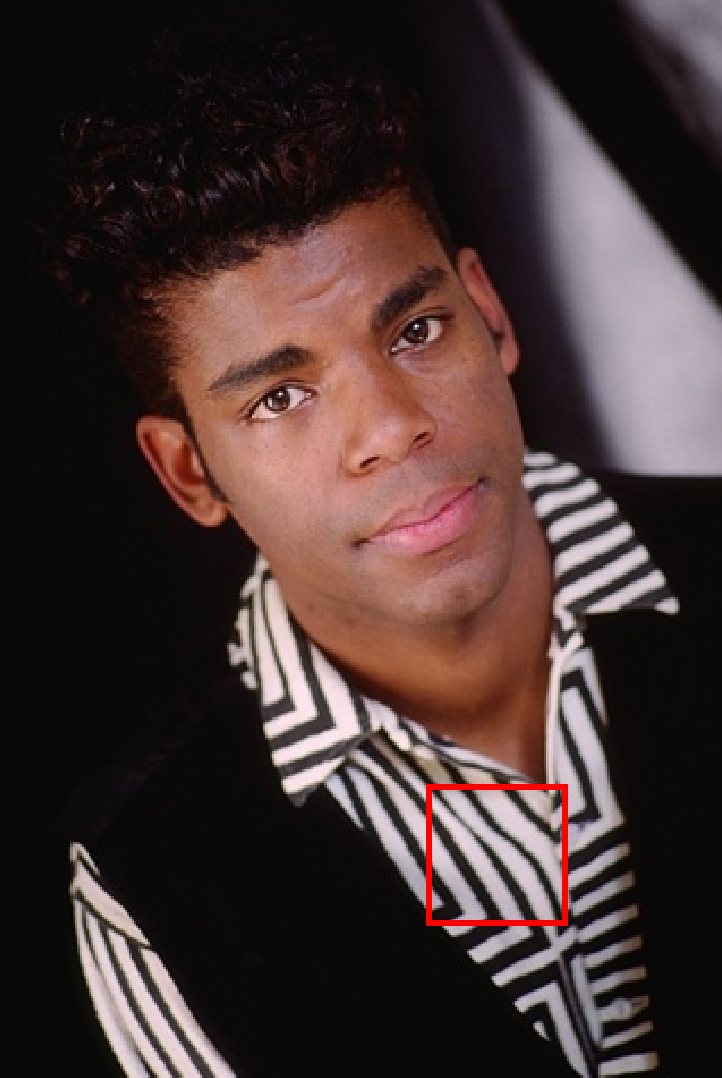}
}
\end{minipage}
\begin{minipage}[t]{.69\textwidth}
\begin{tabular}{ccccc}
\centering
\fcolorbox{red}{yellow}{\includegraphics[height=0.25\textwidth]{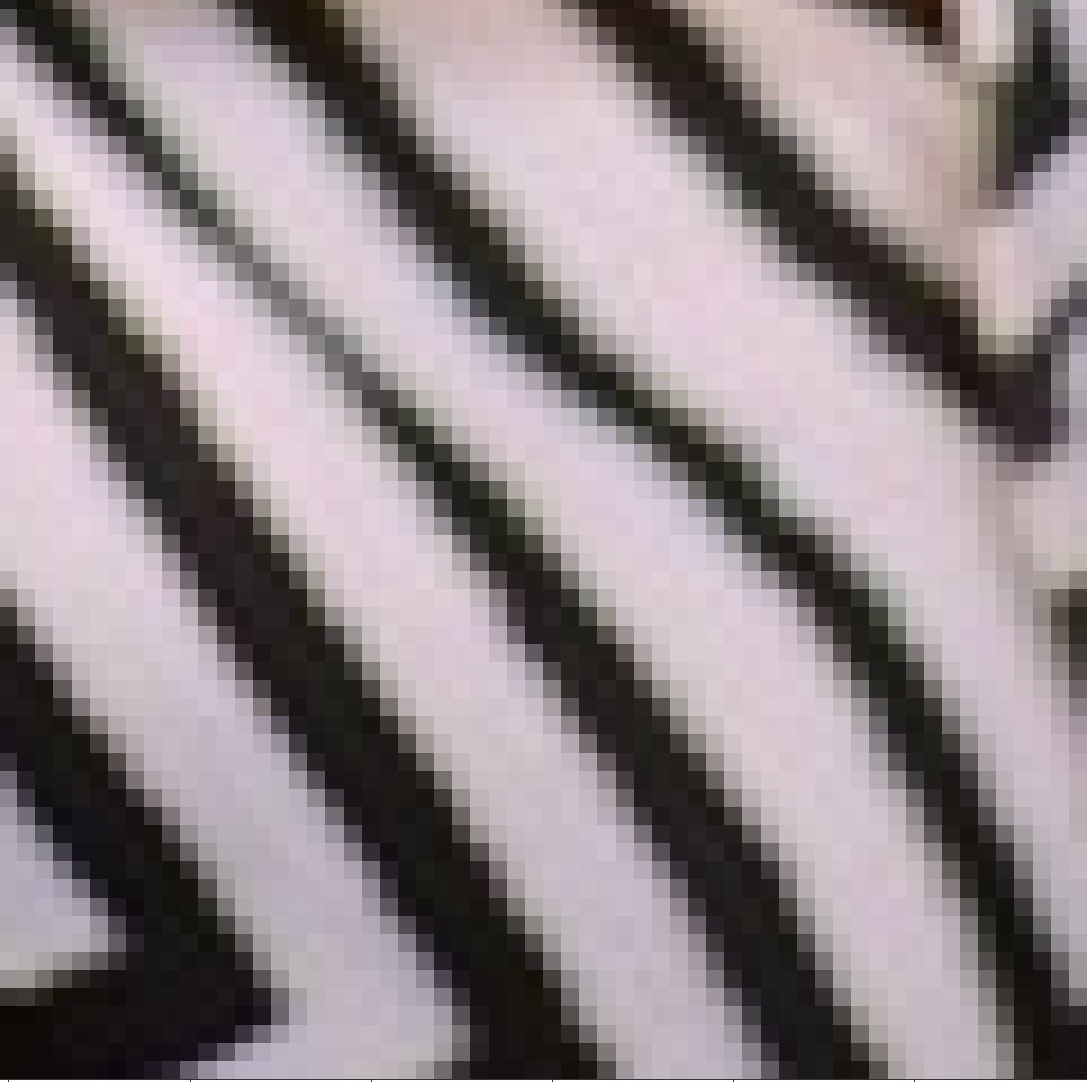}} &
\fcolorbox{red}{yellow}{\includegraphics[width=0.25\textwidth]{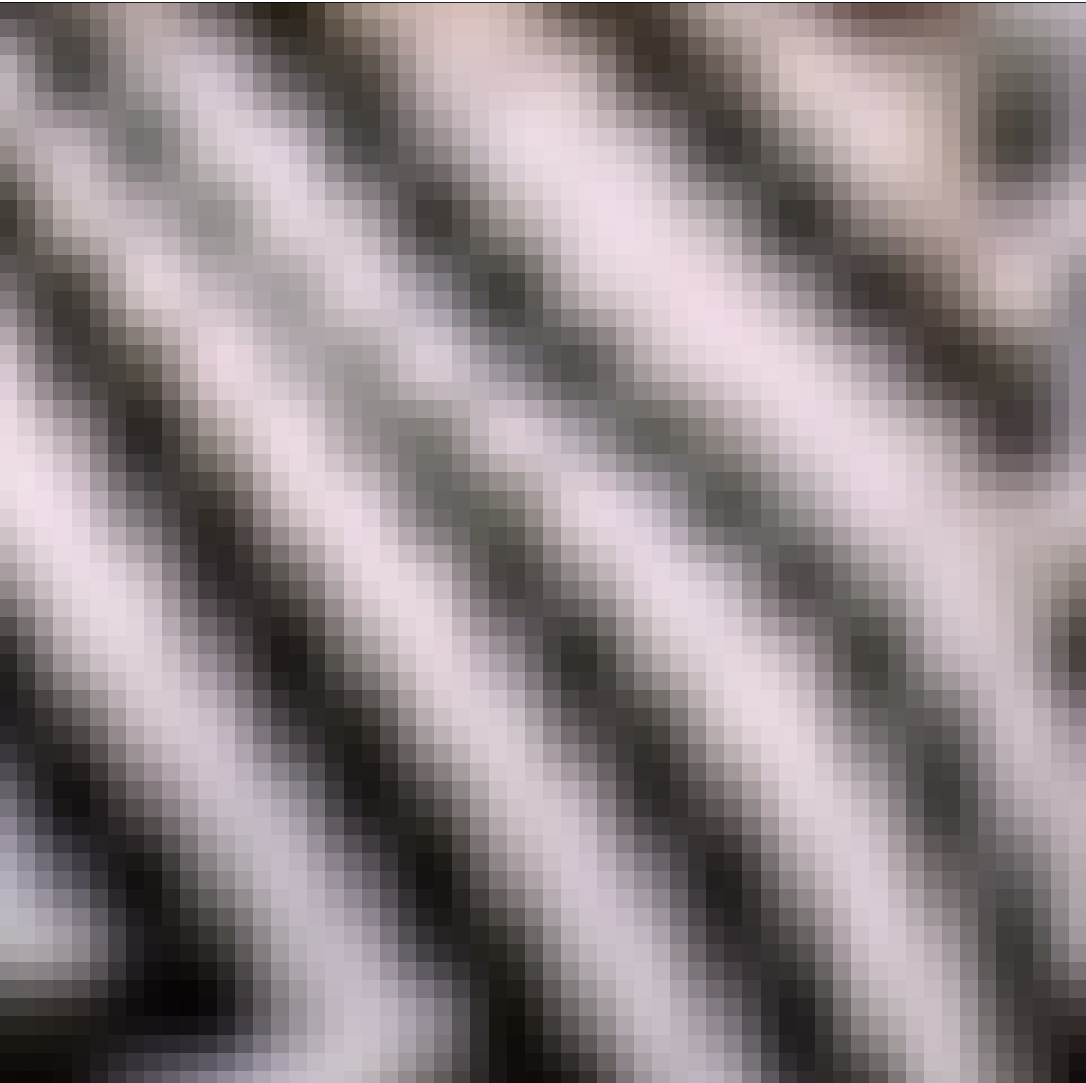}} &
\fcolorbox{red}{yellow}{\includegraphics[width=0.25\textwidth]{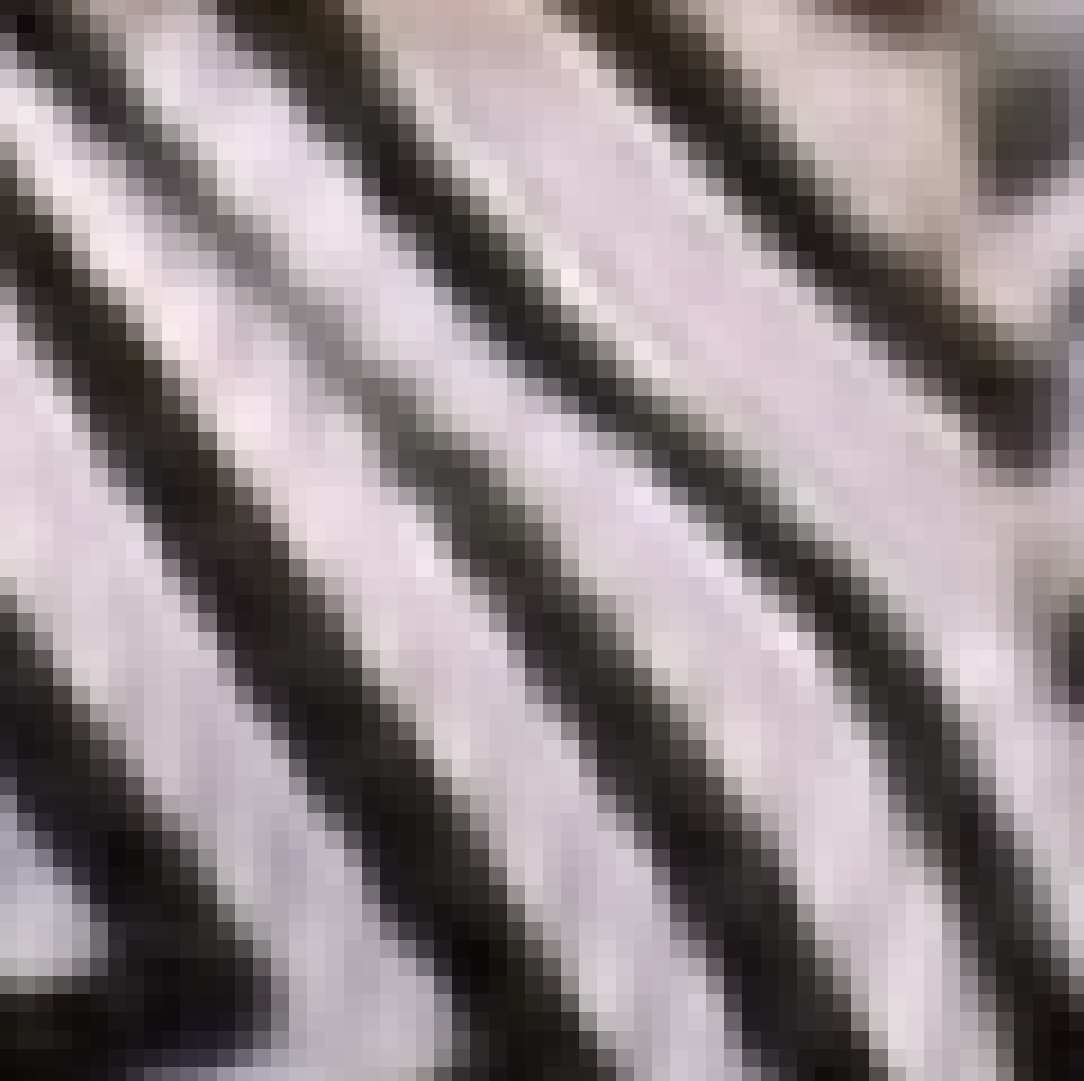}}  &
\fcolorbox{red}{yellow}{\includegraphics[width=0.25\textwidth]{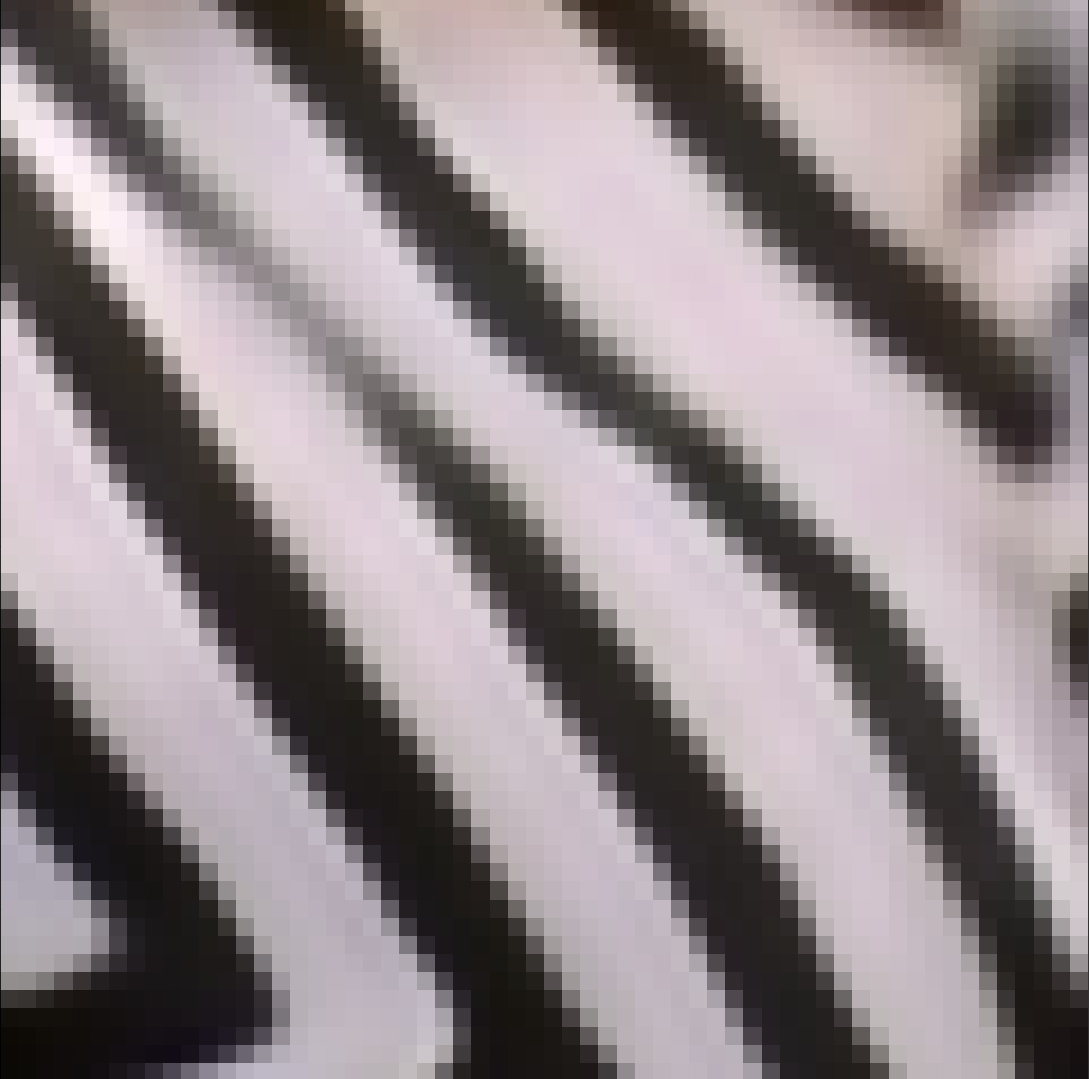}} &
\fcolorbox{red}{yellow}{\includegraphics[width=0.25\textwidth]{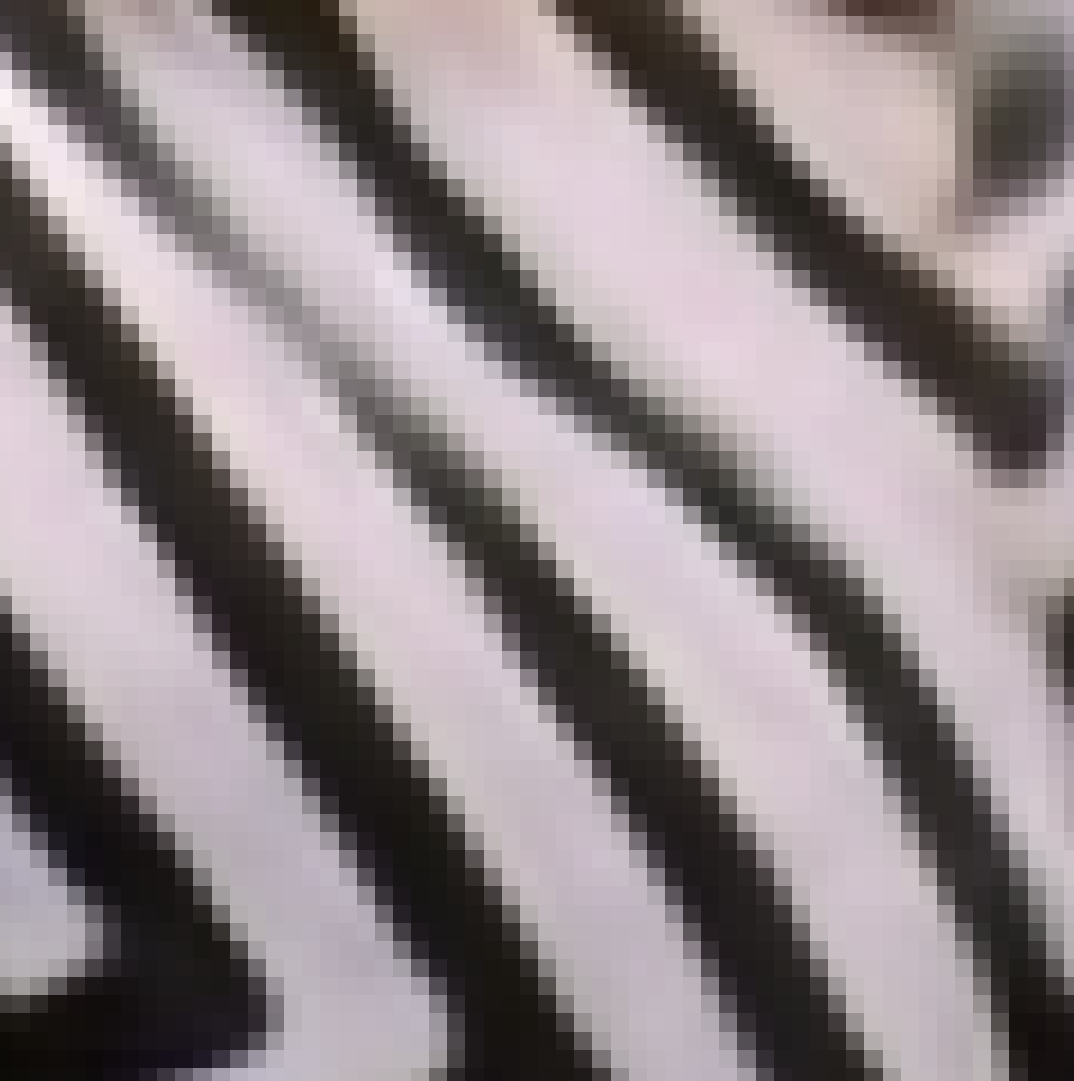}} 
\\
HR &
Bicubic &
SRCNN~\cite{dong2015image}  &
DRCN~\cite{Kim_2016_CVPR1}  &
VDSR~\cite{Kim_2016_CVPR}
\\
&&&&
\\
\fcolorbox{red}{yellow}{\includegraphics[width=0.25\textwidth]{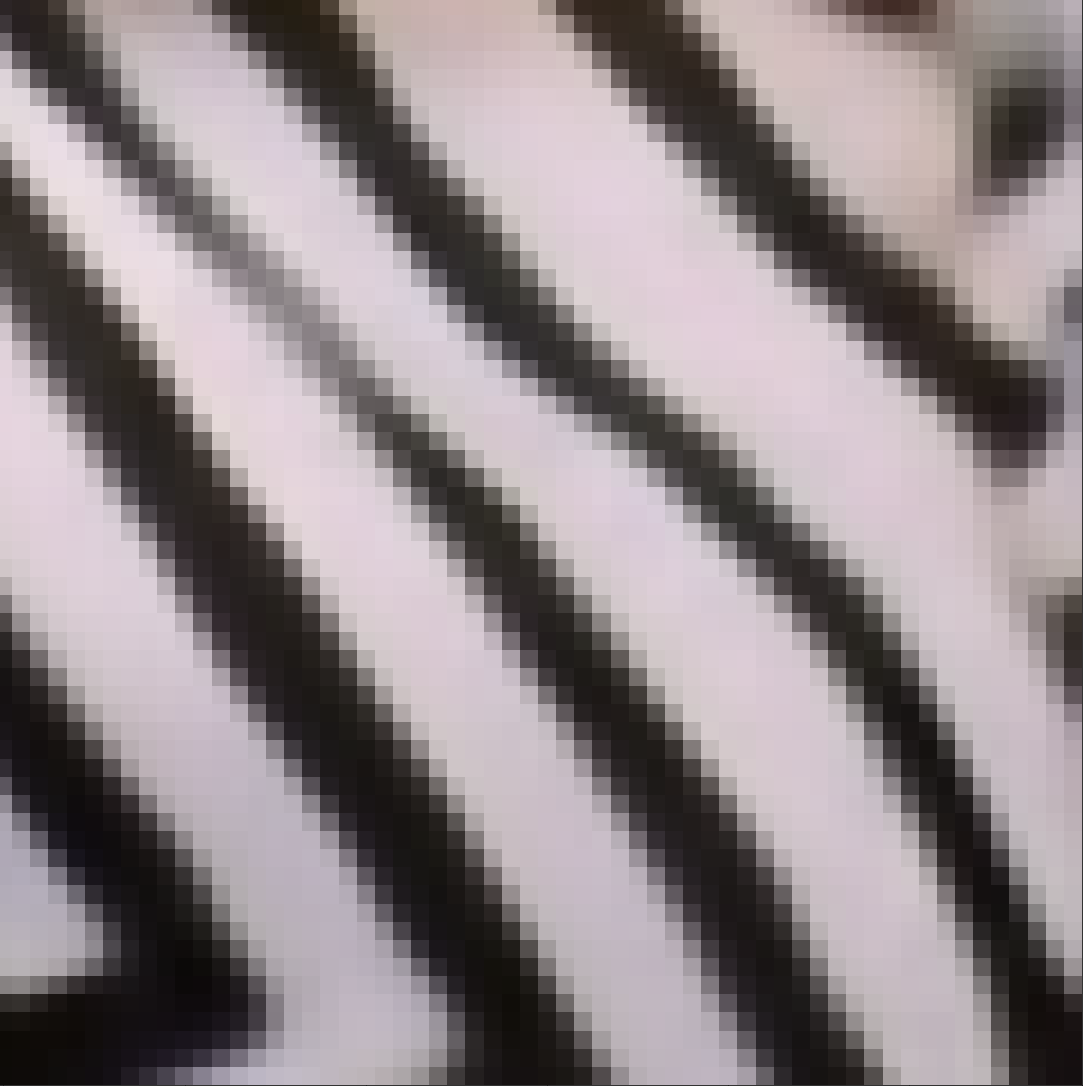}} &
\fcolorbox{red}{yellow}{\includegraphics[width=0.25\textwidth]{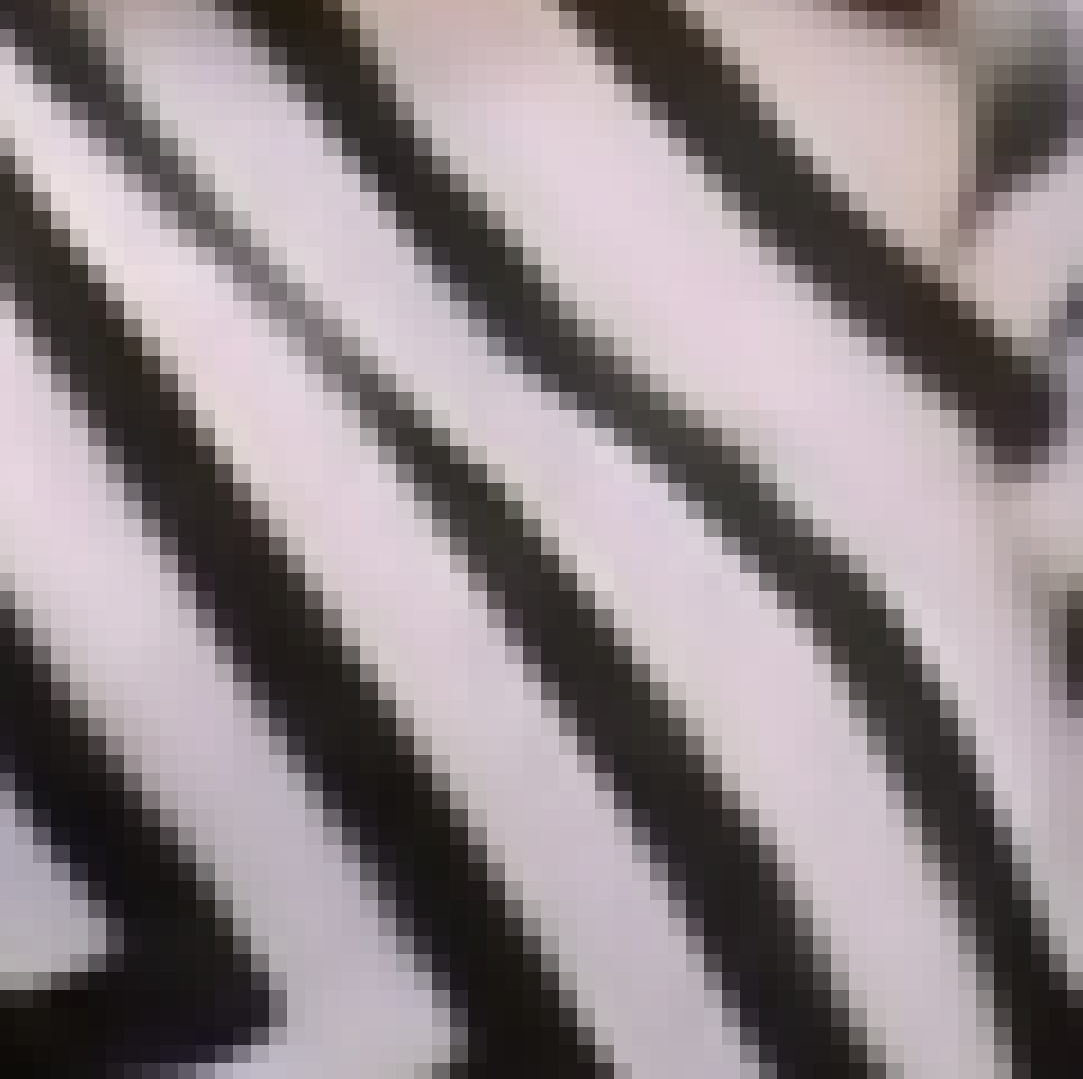}} &
\fcolorbox{red}{yellow}{\includegraphics[width=0.25\textwidth]{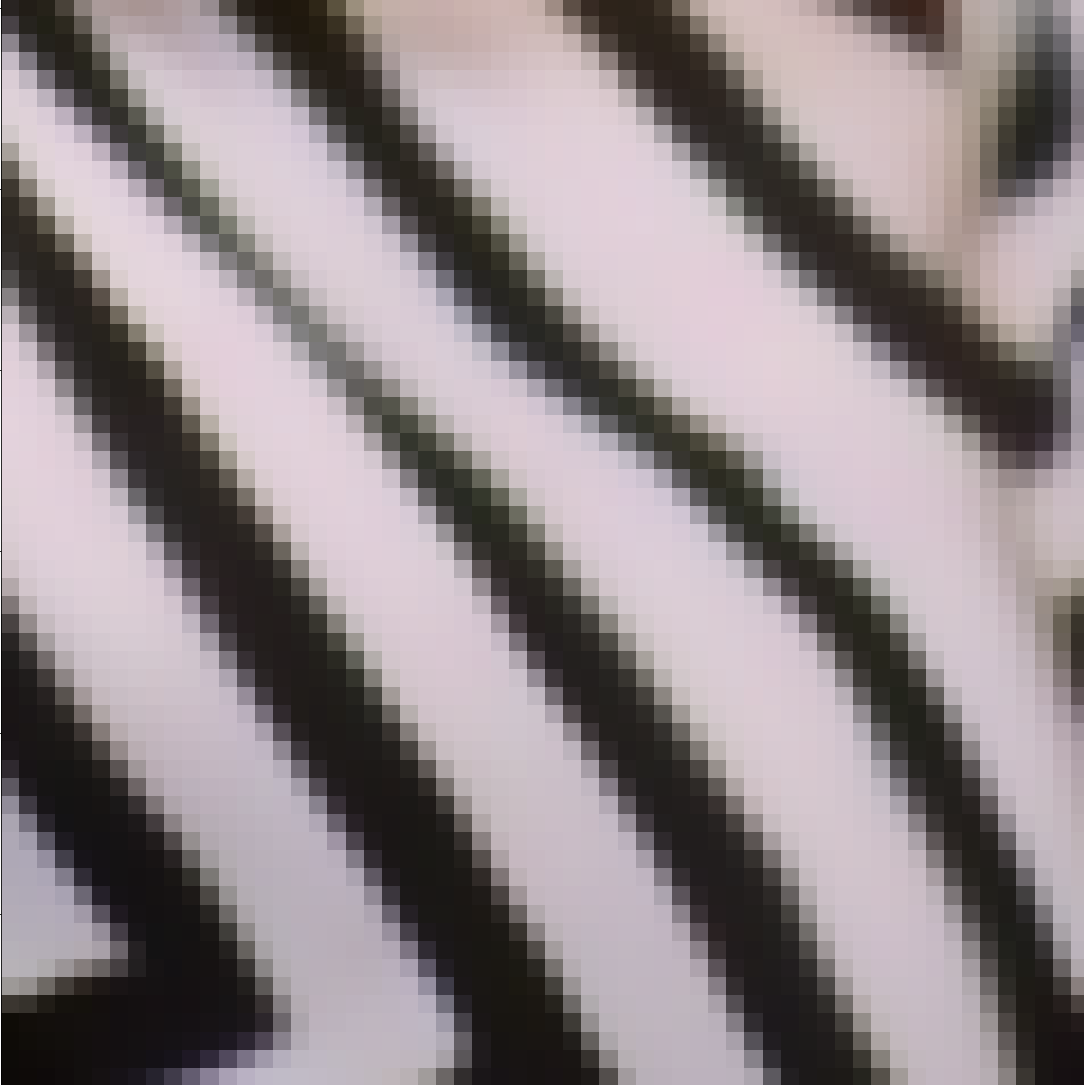}}  & 
\fcolorbox{red}{yellow}{\includegraphics[width=0.25\textwidth]{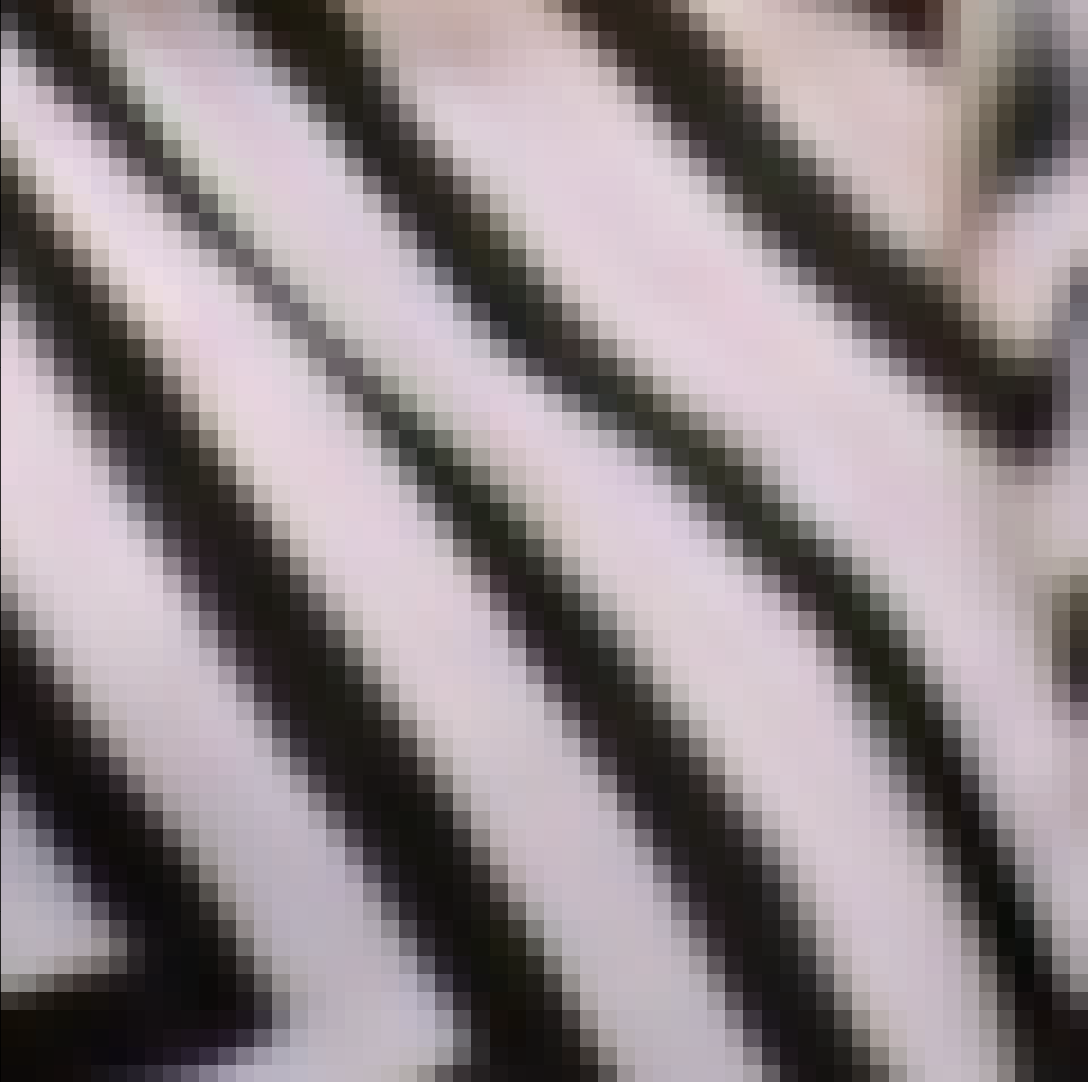}}  &
\fcolorbox{red}{yellow}{\includegraphics[width=0.25\textwidth]{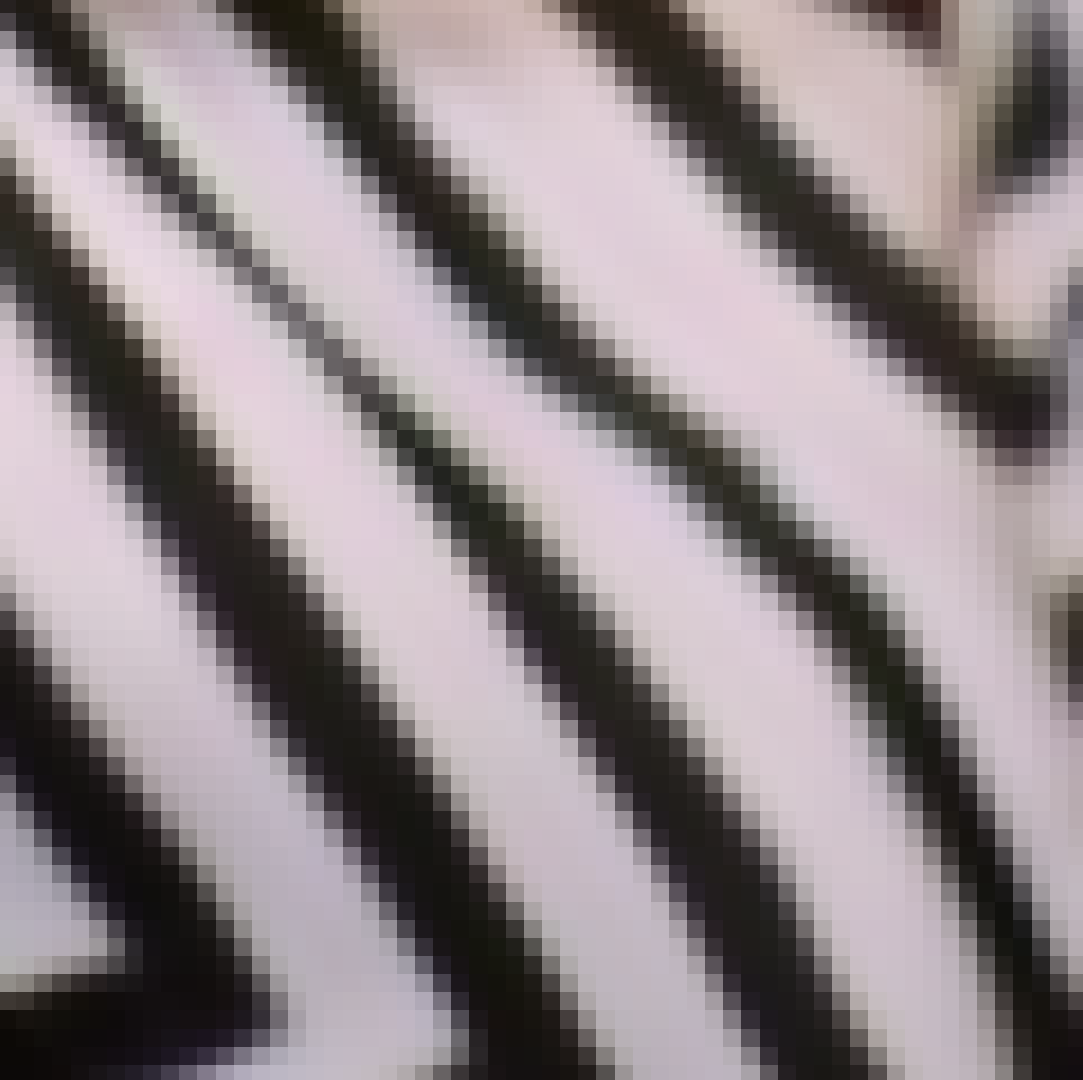}}  
\\
LapSRN~\cite{lai2017deep} &
DRRN~\cite{Tai_2017_CVPR} &
D-DBPN~\cite{Haris_2018_CVPR}  &
RDNLA (ours) &
RDNLA+ (ours)
\end{tabular}
\end{minipage}
}
\caption{Visual results ($4\times$) with model trained on DIV2K~\cite{Timofte_2017_CVPR_Workshops}. The results are presented on image ``$302008$'' from B100~\cite{b100} dataset.}
\label{fig:302008}
\end{figure*}

In Figures~\ref{fig:119082}, \ref{fig:set14_comic}, and ~\ref{fig:302008} we show the visual comparisons on scale $\times4$. For images ``$119082$'' (Figure~\ref{fig:119082}) and ``$302008$'' (Figure~\ref{fig:302008}) we can observe that most of the compared methods generate hallucinated edges (high-frequency information) which is quite common in SISR due to limited information provision and training bias~\cite{Arefin_2020_CVPR_Workshops}. In contrast to the previous methods, RDNLA and RDNLA+ do not generate hallucinated information and thus, provides more reliable results proving effective for super-resolution of remote-sensing data. For image ``005'' (Figure~\ref{fig:set14_comic}) it can be clearly stated that the proposed algorithm retains and generates outputs possessing significantly more high-frequency information compared to the previous methods. The higher-frequency retention is possible due to mask PSNR, the non-local attention mechanism which helps increase the receptive field, and utilization of dense hierarchical features.

\textbf{Study of CM, LRA, and GFB.} The ablation study of the effects of coupled memory (CM), local residual attention (LRA), and global feature blending (GFB) has been depicted in Table~\ref{tab:div2k_arch_changes}. We utilize the DIV2K dataset to observe the effects significantly. 
In all the eight scenarios, there are a total of $\text{P} = 16$ RDNLBs and $\text{D} = 6$ Conv layers, with a growth rate of $\text{G} = 32$. The baseline network obtained without CM, LRA, and GFB (denoted as CM0LRA0GFB0) performs poorly with a PSNR of $36.39$ dB. It indicates that non-overlapping stacking of convolution blocks does not result in a satisfactory output in deep networks. 
\begin{table}[ht!]
    \centering
    \caption{Ablation investigation on the effects of CM, LRA, and GFB. The results are based on Set5~\cite{set5} dataset with a scaling factor $\text{x2}$ in $75$ epochs.}
    \label{tab:div2k_arch_changes}    
    \setlength{\tabcolsep}{0.4\tabcolsep}
    \resizebox{0.5\textwidth}{!}
    {
    \begin{tabular}{|c|c|c|c|c|c|c|c|c|}
    \hline
         & \multicolumn{8}{|c|}{Different combinations of CM, LRA, and GFB} \\ \hline \hline
        CM & \xmark & \cmark & \xmark & \xmark & \cmark & \cmark & \xmark & \cmark\\
        LRA & \xmark & \xmark & \cmark & \xmark & \cmark & \xmark & \cmark & \cmark\\
        GFB & \xmark & \xmark & \xmark & \cmark & \xmark & \cmark & \cmark & \cmark\\ \hline \hline
        \small{PSNR} & \small{$36.39$} & \small{$36.95$} & \small{$36.45$} & \small{$36.61$} & \small{$37.07$} & \small{$37.33$} & \small{$37.06$} & \small{$37.34$}\\ \hline
    \end{tabular}
    }
\end{table}
We then examined how each component contributes towards enhancing the SR output, which resulted in networks CM1LRA0GFB0, CM0LRA1GFB0, and CM0LRA0GFB1 (Col $2-4$ in Table~\ref{tab:div2k_arch_changes}). Every component significantly boosts the model's performance since it contributes to efficient information flow and hierarchical feature extraction. Similar effects can be observed when two and all the three components are utilized in the network (Col $5-8$ in Table~\ref{tab:div2k_arch_changes}), providing further enhanced results with the best $\text{PSNR} = 37.34$ dB (CM1LRA1GFB1). 
It is observed that CM plays a critical role among the three, followed by GFB in image SR.

\begin{figure*}[!htb]
    \centering
    \includegraphics[width=0.9\textwidth]{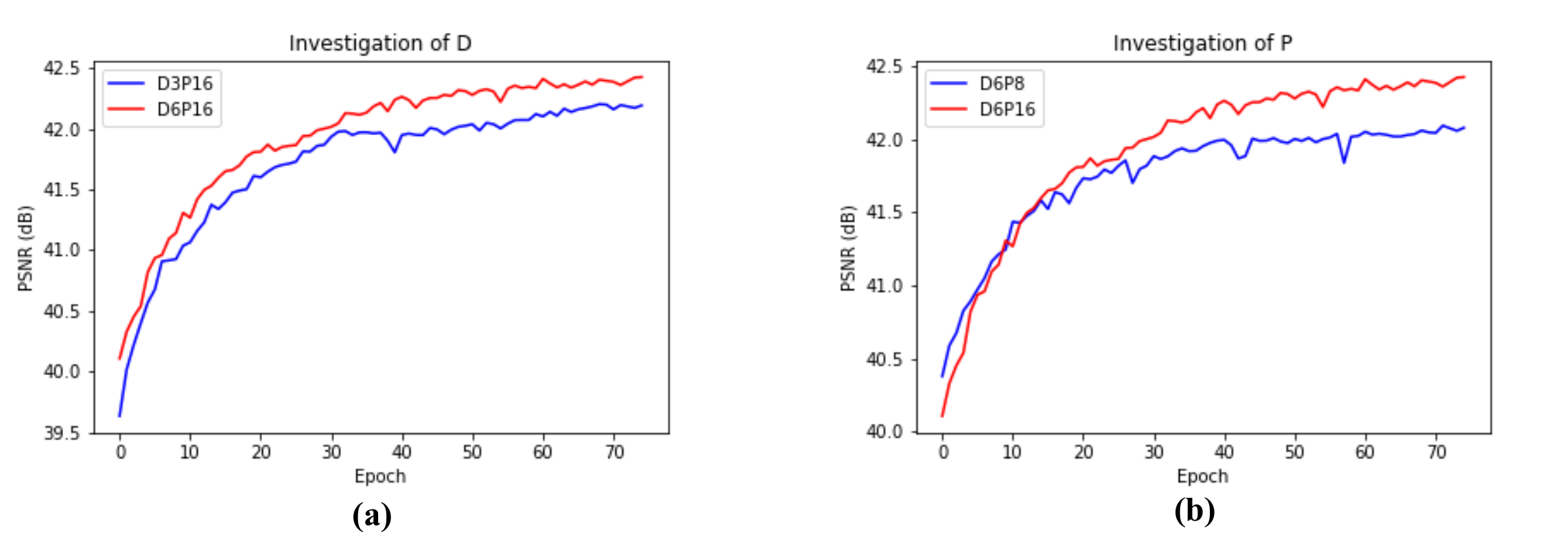}
    \caption{Convergence Analysis of RDNLA with different values of D and P}
    \label{fig:effect_of_PD}
\end{figure*}

\textbf{Study of D and P.} We investigate the basic network parameters: The number of RDNLBs (P) and the number of Conv layers per RDNLB (D). As shown in Figure~\ref{fig:effect_of_PD}, larger P and D would lead to higher performance. This is mainly because a larger P and D increase the depth of the network which allows it to capture long-range dependent hierarchical features for higher performance. With the help of this study, we chose $P=16$ and $D=6$ for all lunar SR experiments. 

\textbf{Effect of Mask Size:} Table~\ref{tab:mask_changes} shows the results of the effect of mask size, $k$, on the SR outputs. 
SR outputs  of all the networks' are reconstructed using an overlap reconstruction mechanism. 
The effective patch we extract during the overlap reconstruction is of dimension $48\times48$. 
However, as observed from Table~\ref{tab:mask_changes}, the best result is obtained at $k = 54$, with the reason being that at $k = 48$, the mask utilized does not provide enough attention to the contextual information at the boundary of the effective patch. 
Therefore, the model does not fully extract information at the boundaries and does not give the highest quality output.

\begin{table}[ht!]
    \centering
    \caption{Effect of mask size, $k$. The results, PSNR/PSNRB, are based on Kaguya DOMs~\cite{haruyama2008global} with a scaling factor $\text{x2}$. Baseline best results are underlined and overall best results are shown in bold.}
    \label{tab:mask_changes}    
    \resizebox{0.5\textwidth}{!}
    {
    \begin{tabular}{|c|c||c|c|}
    \hline
        Loss & $k$ & RDN & \textbf{Proposed}\\\hline \hline
        L1 & $-$ & $69.900/69.824$ & $69.917/69.840$ \\ \hline
        \multirow{4}{*}{Mask-PSNR} & $48$ & $69.916/69.831$ & $69.951/\textbf{69.897}$ \\
         & $54$ & $\underline{69.934}/\underline{69.853}$ & $\textbf{69.971}/69.879$ \\
         & $58$ & $69.921/69.843$ & $69.965/69.889$ \\\hline
    \end{tabular}
    }
\end{table}

The results demonstrate that irrespective of the mask size and architecture, networks trained with mask-PSNR give better SR results than the networks trained with L1 loss. 
Extensive analysis of the test results depicts that not only in terms of average values, but also for each and every reconstructed image, models trained with mask PSNR give at least similar or better outputs (higher PSNR and PSNRB) than models trained with L1 loss.

\begin{figure}[ht!]
    \centering
	\includegraphics[width=0.8\linewidth]{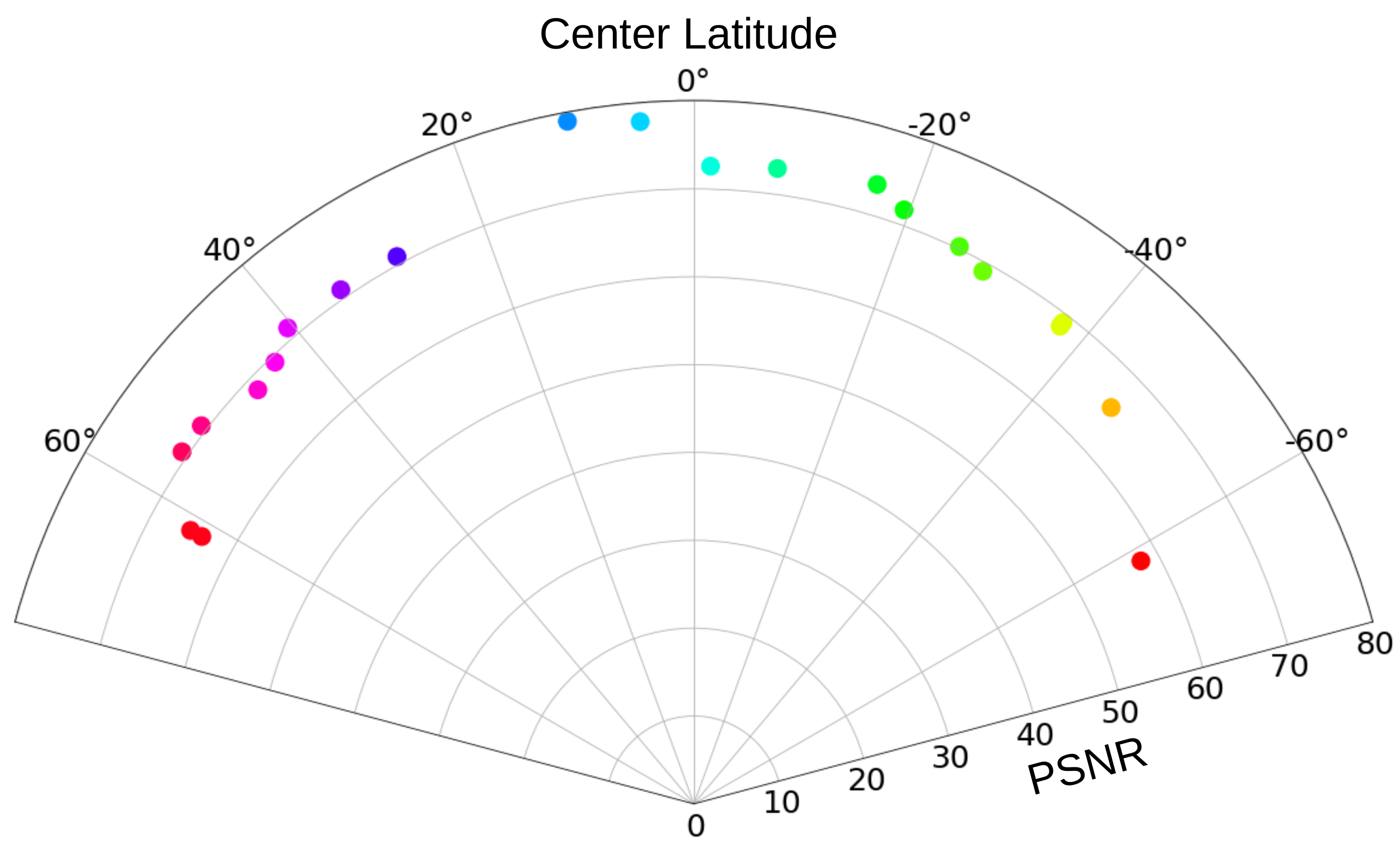}
	\caption{Test result analysis based on the latitude location of the DOMs.}
	\label{fig:lat_psnr_graph}
\end{figure}

\textbf{Effect of Location:} We then plot the Center Latitude (vs) PSNR to evaluate the effect of DOMs' location on the SR output. 
As shown in Figure~\ref{fig:lat_psnr_graph}, it is observed that the images at the equatorial region (around $0^{\circ}$) give significantly enhanced SR outputs compared to near-polar regions. It may be caused by the simple cylindrical projection of the moon's surface. Distortion of shape and scale in a cylindrical projection is minimal in equatorial regions and maximal at the polar regions.

\section{Conclusion}
In this paper, we identify an unresolved issue of data utilization efficiency in satellite missions. We solve it by presenting a simple yet straightforward system design for selective data transmission based on in-orbit inferences. We include an in-orbit SR block in the system design to make better in-orbit inferences. We introduce a novel SR evaluation procedure, an overlap reconstruction mechanism, a new loss function (mask PSNR), and a residual dense non-local attention network for the SR algorithm executability on low-powered device platforms and enhance SR results. Experiments on lunar data showed that the proposed algorithm achieves better PSNR and PSNRB over the baseline residual dense network (RDN). Moreover, our proposed SR algorithm consumes  48\% lesser memory and 67\% lesser power than the RDN, making it more suitable for executing on low-powered devices.

\bibliographystyle{IEEEtran}
\bibliography{rdnla_bib}

\end{document}